\documentclass{bmcart}

\usepackage{natbib}
\usepackage{amsthm,amsmath}
\RequirePackage{natbib}
\RequirePackage{hyperref}
\usepackage[utf8]{inputenc} %unicode support
\usepackage{epstopdf}
\usepackage{graphicx}
\usepackage{longtable}
\usepackage{multirow}
\usepackage{epsfig}
\usepackage{algpseudocode}
\usepackage{algorithm}
\makeatletter
\def\BState{\State\hskip-\ALG@thistlm}
\makeatother
\startlocaldefs
\endlocaldefs

\begin{document}

%%% Start of article front matter
\begin{frontmatter}

\begin{fmbox}
\dochead{Research}


\title{Characterizing Communities of Hashtag Usage on Twitter During the 2020 COVID-19 Pandemic by Multi-view Clustering}

%%%%%%%%%%%%%%%%%%%%%%%%%%%%%%%%%%%%%%%%%%%%%%
%%                                          %%
%% Enter the authors here                   %%
%%                                          %%
%% Specify information, if available,       %%
%% in the form:                             %%
%%   <key>={<id1>,<id2>}                    %%
%%   <key>=                                 %%
%% Comment or delete the keys which are     %%
%% not used. Repeat \author command as much %%
%% as required.                             %%
%%                                          %%
%%%%%%%%%%%%%%%%%%%%%%%%%%%%%%%%%%%%%%%%%%%%%%

\author[
   addressref={aff1},                   % id's of addresses, e.g. {aff1,aff2}
   %corref={aff1},                       % id of corresponding address, if any
   %noteref={n1},                        % id's of article notes, if any
   email={icruicks@andrew.cmu.edu}   % email address
]{\inits{IC}\fnm{Iain J} \snm{Cruickshank}}
\author[
   addressref={aff1},
   email={kathleen.carley@cs.cmu.edu}
]{\inits{KC}\fnm{Kathleen M} \snm{Carley}}

%%%%%%%%%%%%%%%%%%%%%%%%%%%%%%%%%%%%%%%%%%%%%%
%%                                          %%
%% Enter the authors' addresses here        %%
%%                                          %%
%% Repeat \address commands as much as      %%
%% required.                                %%
%%                                          %%
%%%%%%%%%%%%%%%%%%%%%%%%%%%%%%%%%%%%%%%%%%%%%%

\address[id=aff1]{%                           % unique id
  \orgname{CASOS, Carnegie Mellon University}, % university, etc
  \street{5000 Forbes Ave},                     %
  %\postcode{}                                % post or zip code
  \city{Pittsburgh, PA},                              % city
  \cny{USA}                                    % country
}

%%%%%%%%%%%%%%%%%%%%%%%%%%%%%%%%%%%%%%%%%%%%%%
%%                                          %%
%% Enter short notes here                   %%
%%                                          %%
%% Short notes will be after addresses      %%
%% on first page.                           %%
%%                                          %%
%%%%%%%%%%%%%%%%%%%%%%%%%%%%%%%%%%%%%%%%%%%%%%

%\note{Sample of title note}     % note to the article
%\note[id=n1]{Equal contributor} % note, connected to author

\end{fmbox}% comment this for two column layout

%%%%%%%%%%%%%%%%%%%%%%%%%%%%%%%%%%%%%%%%%%%%%%
%%                                          %%
%% The Abstract begins here                 %%
%%                                          %%
%% Please refer to the Instructions for     %%
%% authors on http://www.biomedcentral.com  %%
%% and include the section headings         %%
%% accordingly for your article type.       %%
%%                                          %%
%%%%%%%%%%%%%%%%%%%%%%%%%%%%%%%%%%%%%%%%%%%%%%

\begin{abstractbox}

\begin{abstract} % abstract
The COVID-19 pandemic has produced a flurry of online activity on social media sites. As such, analysis of social media data during the COVID-19 pandemic can produce unique insights into discussion topics and how those topics evolve over the course of the pandemic. In this study, we propose analyzing discussion topics on Twitter by clustering hashtags. In order to obtain high quality clusters of the Twitter hashtags, we also propose a novel multi-view clustering technique which incorporates multiple different data types that can be used to describe how users interact with hashtags. The results of our multi-view clustering show that there are distinct temporal and topical trends present within COVID-19 twitter discussion. In particular, we find that some topical clusters of hashtags shift over the course of the pandemic, while others are persistent throughout, and that there are distinct temporal trends in hashtag usage. This study is the first to use multi-view clustering to analyze hashtags and the first analysis of the greater trends of discussion occurring online during the COVID-19 pandemic.
\end{abstract}

%%%%%%%%%%%%%%%%%%%%%%%%%%%%%%%%%%%%%%%%%%%%%%
%%                                          %%
%% The keywords begin here                  %%
%%                                          %%
%% Put each keyword in separate \kwd{}.     %%
%%                                          %%
%%%%%%%%%%%%%%%%%%%%%%%%%%%%%%%%%%%%%%%%%%%%%%

\begin{keyword}
\kwd{Social Media}
\kwd{Clustering}
\kwd{Multi-view Data}
\kwd{COVID-19}
\end{keyword}

% MSC classifications codes, if any
%\begin{keyword}[class=AMS]
%\kwd[Primary ]{}
%\kwd{}
%\kwd[; secondary ]{}
%\end{keyword}

\end{abstractbox}
%
%\end{fmbox}% uncomment this for twcolumn layout

\end{frontmatter}

%%%%%%%%%%%%%%%%%%%%%%%%%%%%%%%%%%%%%%%%%%%%%%
%%                                          %%
%% The Main Body begins here                %%
%%                                          %%
%% Please refer to the instructions for     %%
%% authors on:                              %%
%% http://www.biomedcentral.com/info/authors%%
%% and include the section headings         %%
%% accordingly for your article type.       %%
%%                                          %%
%% See the Results and Discussion section   %%
%% for details on how to create sub-sections%%
%%                                          %%
%% use \cite{...} to cite references        %%
%%  \cite{koon} and                         %%
%%  \cite{oreg,khar,zvai,xjon,schn,pond}    %%
%%  \nocite{smith,marg,hunn,advi,koha,mouse}%%
%%                                          %%
%%%%%%%%%%%%%%%%%%%%%%%%%%%%%%%%%%%%%%%%%%%%%%

%%%%%%%%%%%%%%%%%%%%%%%%% start of article main body
% <put your article body there>

%%%%%%%%%%%%%%%%
%% Background %%
%%

\section*{Introduction}

At the time of the writing of these words, the world is undergoing a pandemic. This pandemic, which is caused by the SARS-CoV-2 virus and often referred to as the COVID-19 pandemic, has caused immense societal and economic disruption across the world. Since the onset of the COVID-19 pandemic many nations have adopted a social-distancing strategy which has had the unintended consequence of emphasizing and increasing the role of social media in linking people together \cite{article19:2020}, \cite{Hussain:2020}. Consequently, the study of social media data during the current COVID-19 pandemic can provide unique insights into online social behavior.

Thus far, much of the work with COVID-19 social media data has focused on the prevalence and spread of  COVID-19 misinformation. There has been less work on understanding what are the important topics of discussion associated with the pandemic and how those discussion topics may change over the course of the pandemic. One social media innovation which can be used to characterize and understand topics of social media conversations are hashtags. Hashtags are a social media innovation which were designed to allow users to easily find and interact with certain discussion topics. So, in this work, we propose clustering hashtags from COVID-19 social media data in order to understand the topics of discussion happening within the greater COVID-19 social media discussion.

In order to cluster hashtags from social media data, we propose the use of multi-view clustering. Many naturally-occurring, social phenomena give rise to multiple types and views of data. So, using a clustering method that can exploit information from all of those views should result in a better clustering of the data. To date and the best of the authors' knowledge, most clustering of social media data is not clustered by multi-view clustering. So, in this work, we propose the use of multi-view clustering in order to better understand the discussion topics surrounding COVID-19 on social media. In particular, we analyze topical clusters of twitter data by multi-view clustering on hashtags, where we use the co-occurrence of hashtags, tweet text, twitter users, and Uniform Resource Locators (URLs) that co-occur with hashtags in a tweet as the views to cluster the hashtags. The main contributions of this work are summarized as follows:
\begin{itemize}
    \item The first use of a multi-view clustering technique and approach to understand topical groups of hashtags.
    \item The first use of a multi-view clustering technique on a large, social-based data set; multiple collections each consisting of upwards of 85,000 objects are clustered in this study whereas most multi-view clustering techniques have been used on data sets, social-based or otherwise, of at most one collection of 70,000 objects (i.e. full MNIST data set).
    \item Characterization of topical clusters of hashtags that give distinct insight into what conversations surround the COVID-19 pandemic on twitter, such as co-opting of the calamity to support different causes and a persistent coupling of U.S. politics related hashtags with conspiracy theory related hashtags.
\end{itemize}

The article is organized as follows: In the next section we provide some background on COVID-19 social media analyses and multi-view clustering. In the data section, we provide an overview of the data used in this study and the steps taken to process the data. In the methodology section we describe the methodology proposed in this work for clustering multi-view, social-based data, like hashtags, and the settings used for this study's data. In the results section we present the results of the multi-view clustering, identify different time periods of hashtag usage, perform an ensemble clustering of the hashtag clusters to produce prototypical hashtags clusters for the different time periods, and then qualitatively analyze those clusters and the user bases of the clusters. Finally, we conclude the study with a discussion of the results and directions for future work.

\section*{Background}

Current study into online social behavior during the COVID-19 pandemic has largely focused on how information and misinformation operates during a pandemic. This is because good information is a key enabler to combat the effects of the pandemic whereas misinformation can exacerbate its effects \cite{Hussain:2020}, \cite{Gallotti:2020}. Recent studies into the prevalence and persistence of misinformation have shown that misinformation on the COVID-19 pandemic has been especially persistent and spreads through online social networks quickly \cite{Yang:2020}, \cite{article19:2020}, \cite{Boberg:2020}. The spread of COVID-19 misinformation has become so problematic and widespread that many many researchers are referring to it as an `Infodemic' \cite{Gallotti:2020}, \cite{Cinelli:2020}, \cite{article19:2020}. The Infodemic is characterized by a virus-like spread of misinformation across many different communication mediums, most notably online social networks. Additionally, other researchers have identified important mechanisms by which the misinformation propagates in social media. Recent research has identified the importance of bots in the spread of misinformation \cite{Ferrara:2020}. Other research has highlighted the role of alternative news sources and user characteristics like political beliefs in the spread of COVID-19 misinformation \cite{Boberg:2020}, \cite{Huang:2020}.

One area that is less clear is how social media users may be changing their behavior and how social media communities and discussions are changing in response to the COVID-19 pandemic and Infodemic. While the aforementioned research has shown that social media users are spreading COVID-19 misinformation, sometimes even faster than good information, its not clear how users' interactions or how discussion communities may be changing during the pandemic. It is also not clear if there are any topical areas of focus for social media users during the ongoing pandemic. For example, are social media discussions focusing around topics like health and welfare or around politics and business, or any combination thereof?

One of the recent social media innovations that have been used to track and understand conversations and conversational topics are hashtags. Hashtags originated in 2007 on the social media platform Twitter as a means of allowing users to efficiently retrieve information relevant to a topic \cite{Zhang_twitter:2019}. The use of hashtags on Twitter has expanded to not only be a means of characterizing discussion topics, but also a means of predicting user links and characterizing both communities of users as well as the users themselves \cite{Zhang_twitter:2019}, \cite{Xiao:2014}, \cite{Shapp:2014}, \cite{Saxton:2015}, \cite{Sheldon:2019}. As such, clustering of hashtags can be used to understand topics of interest for social media users and the communities that form around certain discussions \cite{Kywe:2012}, \cite{Vicient:2015}. The clustering of hashtags has been done by either the text context used with the hashtags, co-occurrence of hashtags within the same social media post (i.e. same tweet), or by having similar users that use the same hashtags \cite{Kywe:2012}, \cite{Xiao:2014}. More recent work on clustering hashtags has focused on better feature engineering of the text that accompanies hashtags in order to capture the semantic meaning of the text and thus better hashtag clusters \cite{Vicient:2015}. To date, no work has attempted to combine all of these different views of hashtag usage and use multi-view clustering to cluster hashtags in order analyze social media data for understanding topics of interest for social media users. 

\subsection*{Multi-view Clustering}

Multi-view clustering techniques are techniques designed to handle clustering of objects which can be described by more than one data source. Many different real-world, social phenomena give rise to `views' of data which are often different types of data that can be used to describe the same set of actors. For example, social media users can post content, which could give rise to a text view, and have interactions with each other, which can give rise to network views. So, multi-view clustering aims to fuse the information from these different views of the data to produce one clustering of the object that created the data \cite{Morency:2019}, \cite{Ye:2018}, \cite{Yang:2018}, \cite{Baltrusaitis:2017}. There have been a surge of new techniques developed in multi-view clustering for handling genetic \cite{Zitnik:2018}, \cite{Huang:2017} and image data \cite{Yang:2018}, \cite{Bai:2018}. Many of these techniques rely on producing graphs of each view of the data and then collectively clustering those graphs \cite{Morency:2019}, \cite{Baltrusaitis:2017}.  By doing so, these techniques can then define one function across all of the views for measuring the goodness of the clustering and exploit properties of graphs in modeling and preserving complex structures in the data, as is often done with spectral clustering \cite{Guo:2014}, \cite{Xia:2014}, \cite{Xu:2016}, \cite{Yu:2017}. While these techniques can be used --- at least in principle --- with any kind of data, very few have been applied to social-based, multi-view data. 

Part of the difficulty in multi-view clustering of social media data like hashtags are that the data is often very large (on the order of tens or hundreds of thousands of hashtags being used in any given conversation) which can pose problems for many of the existing multi-view clustering techniques. Additionally, social media data often have partially-complete views of data; social-media users and objects, like hashtags, may not have any interactions within some views. For example, a hashtag may never co-occur with another hashtag, or a user may never engage in an activity like re-tweeting. These partially complete views pose challenges for many existing multi-view clustering techniques as these objects naturally become isolates or small connected components in the view graphs used in the clustering.

Finally, Within the realm of social-based data, and in particular social network analysis, there are two main areas of research that deal with multi-view data. The first is multi-layer or multiplex social network analysis. Multiplex networks are networks in which a node has more than one type of link connecting it to other nodes \cite{Aleta:2019}. These networks can be modeled as a collection of networks that are defined over the same nodeset but have different links within each of the networks. So, in this data format, each network represents a possible view of the data. Multi-layer networks contain the same networks as multiplex networks, but with the addition of inter-layer links where a node in one layer can be connected to nodes in other layers \cite{Aleta:2019}. As with multiplex networks, each layer can be considered a view of the data. There have been a host of techniques designed to work with multi-layer and multiplex networks, most of which leverage Network Modularity as the objective function to cluster the networks \cite{Mucha:2010}, \cite{Pamfil:2018}, \cite{Tagarelli:2017}, or use a stochastic block-model \cite{Jingchao:2015}, \cite{Pamfil:2018}, \cite{Yuming:2019}. So, while there are a wealth of techniques available, these techniques are limited to multi-view data that only consists of networks.

The second main area of multi-view clustering of social-based data are attributed networks. Attributed networks are networks which also have additional information on the nodes \cite{Chunaev:2019}. So, an attributed network will have two views of data; one view which is the network itself and a second view of features describing the nodes present in the network. Attributed networks are often clustered by either combining the attribute information into the network itself and then clustering that network by standard network clustering techniques \cite{Chunaev:2019}, \cite{Papadopoulos:2015}, \cite{Alinezhad:2019}, \cite{Papadopoulos:2017}, or by defining a new measure of network modularity that incorporates a term for the attributes \cite{Combe:2015}, \cite{Chunaev:2019}. So, attributed network clustering is limited to just the multi-view data scenario in which there is one network view and one non-network view of the data.

\section*{Data}

The data for this analysis comes from Twitter's streaming API \footnote{https://developer.twitter.com/en/docs/tweets/filter-realtime/guides/basic-stream-parameters}. The data collection was done using a list of key words including “coronavirus”, “coronaravirus”,
“wuhan virus”, “wuhanvirus”, “2019nCoV”, “NCoV”, “NCoV2019” \cite{Huang:2020}. The collected data spans the time period from 1 February 2020 to 30 April 2020 and consists of over 300 million tweets that have, on average, 45,000 unique hashtags per day. The following figure, Figure \ref{fig:daily_unique_hashtag_counts.png} depicts the daily statistics concerning the use of hashtags within the data set. It should be noted that as a prepossessing step, any hashtag that was not used in at least 3 tweets was not included in the data. These hashtags are often misspellings of more widely used hashtags.

\begin{figure}[ht]
    \centering
    \includegraphics[width=0.9\textwidth]{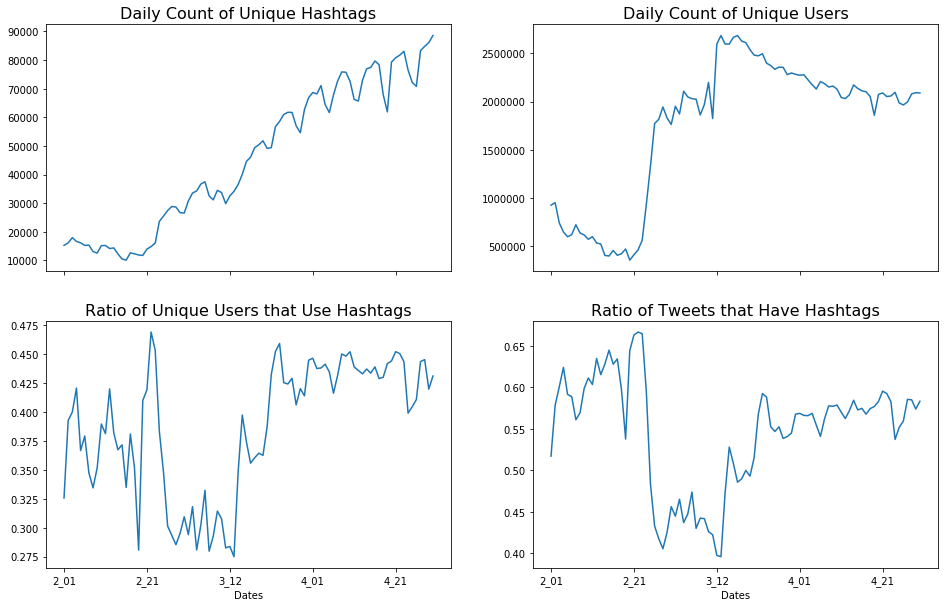}
    \caption{Daily Statistics of the COVID-19 Twitter Data from 1 February 2020 to 30 April 2020. Use of Hashtags by users and within tweets remains high and persistent over the time period.}
    \label{fig:daily_unique_hashtag_counts.png}
\end{figure}

Hashtag usage within the data is both prevalent and increases in the diversity of hashtags being used over time. The use of unique hashtags generally increase over the time period of data collection and displays some weekly cyclical patterns as well (i.e. slight drops in the number of unique hashtags being used on weekend days). It is interesting to note that these counts are counts of unique hashtags and not the total use of hashtags. So, it is possible that as the scope of the COVID-19 pandemic expanded, hashtags that were originally unassociated with the COVID-19 pandemic end up becoming a part of the conversation. It is also possible that as the scope of the pandemic expanded, that new hashtags were invented to better address the changing needs of the conversation about the pandemic. Additionally, there is a relatively high ratio of hashtags being used in tweets throughout all of the data (greater than 40\% of tweets have a hashtag). While this is in part due to the means of collecting the data, it also reflects a trend observed by other authors of increasing hashtag usage among social media users generally \cite{Sheldon:2019}, \cite{Zhang_twitter:2019}. The ratio of hashtag usage in tweets has three observable phases over time. In the first phase, from 1 February to 24 February, the ratio of hashtags present in tweets remains at its highest, with a slight positive trend. Then, starting on the 25th of February the ratio of hashtags in tweets decreases and remains lower until the 15th of March. Finally, in the third phase from the 16th of March until the 30th of April, the ratio of hashtags being used in tweets remains stable at around 55\% of the tweets. So, while there is a high percentage of tweets that have at least one hashtag, this percentage is not stable over time. While its not clear why this temporal trend exists, it is likely related to the dynamic nature of the users interacting on Twitter over the various stages of the COVID-19 pandemic.

There is also a reasonably high percentage of hashtag use among individual users, which generally increases with time. As with the ratio of tweets that feature a hashtag, the ratio of users that use at least one hashtag on a given day breaks into three observable periods. The first period is a period of a lot of oscillation in daily hashtag usage centered around 37\%. The second period is a decline in hashtag use corresponding to the same period of decreased hashtags being used in individual tweets, from 25 February to 15 March. The third period is from the 16th of March to the 30th of April and has a sustained hashtag use at around 43\% of users using at least one hashtag in a tweet per day. Overall, 50.47\% of users use at least one hashtag during the three months that data was collected. In terms of individual hashtag usage, for users in general the usage statistics of hashtags are as follows: min: 0, max: 100\%, mean: 31.6\%, and standard deviation: 38.8\% of their tweets featuring hashtags. Of those users that use at least on hashtag in their tweets, the usage statistics become: min: 1.9\%, max: 100\%, mean 62.6\%, and standard deviation: 32.4\% of their tweets featuring hashtags.

This hashtag usage takes place in a background of variable trends in the number of unique daily users present within the data set. There is a declining number of unique users within the data from the 1st of February to around the 21st of February, at which point there is a large increase in the number of unique users until the 20th of March. From the 20th of March to the end of April, the number of daily unique users begins to decline again. So, while the use of hashtags increases as well as the use of unique hashtags increases, the number of unique users actually declines. So, during the early stages of the pandemic there are a fairly small number of users tweeting relevant tweets which often have hashtags which then transitions over the course of the pandemic to a much larger user base that does not initially use many hashtags. This observation suggests that user's interactions on Twitter are dynamic over the course of a pandemic and that discussion topics, in the form of hashtags are also dynamic over the course of a pandemic. It also suggests that hashtag usage becomes more widely adopted after an initial surge in users possibly as means of better characterizing the new and burgeoning conversations happening on Twitter surrounding the pandemic. All together, the nature of the use of hashtags and the hashtags in use have likely changed over the course of the COVID-19 pandemic.

\subsection*{Data Processing}

The aforementioned collection of raw tweets was then transformed into view data for clustering of the hashtags. First, the data was separated into days. Second, the daily tweet data was preprocessed as follows: First, any hashtag which was used in less than 3 tweets was excluded from the data. Second, the tweet text was preprocessed by removing all of hashtags, URLs, symbols (i.e. emojis, punctuation, etc.), and twitter-specific tags (i.e. mentions, quotes, etc.) from the tweet text. From there, for each hashtag, the accompanying text, other hashtags, the user, and any URLs used in each tweet that contained that hashtag were extracted. This data was then separated into four separate views of the data. The first view is the text view and contains all of the pre-processed text from each tweet that features a hashtag. The intuition behind this view is that the text accompanying a particular hashtag may give insight into how and what a particular hashtag is used for in discussions. The accompanying tweet text, whether used in a raw form or given semantic enhancement has been previously used to cluster hashtags \cite{Kywe:2012}, \cite{Vicient:2015}. The second view is the users which tweet a hashtag with the idea that users may be partial to tweeting particular hashtags as part of a discussion. Shared users have also been used to cluster hashtags in previous works \cite{Kywe:2012}. The third view is the URLs which co-occur with a hashtag. Since URLs are often used as information to support claims in tweets, this view should give insight into what information is underlying the use of certain hashtags. Finally, the fourth view is the co-occurrence of hashtags within tweets. Within any given tweet using hashtags, a user may use multiple, related hashtags as part of their tweet. The co-occurrence of hashtags is frequently used to create hashtag-to-hashtag networks for analysis by standard network science techniques \cite{Vicient:2015}. So, all together, the collected tweets are processed to create four different views of the data for each day in order to cluster the hashtags.

\section*{Methodology}

In this section, the method for performing a multi-view clustering of the hashtags is detailed. As was mentioned in the background section, multi-view clustering of social-based objects poses some distinct challenges. Namely, there are hundreds of thousands of hashtags that need to be clustered, partially complete views, and the views consist of both network and non-network data. As such, we propose a new method of multi-view clustering that can handle all of the aspects of this data. As with any multi-view clustering technique, there are two main requirements: conversion of the views of the data into a format which can fuse information from the views together, and a clustering goodness function that is defined for all of the views used in the clustering. For the first requirement, we adopt the method used in many multi-view clustering techniques of converting all of the views of the data into graphs. This allows for preservation of local structures in the data and provides a flexible format that can be used with any data type \cite{Qiao:2018}, \cite{Brugere:2018}. For the second requirement, we use network modularity due to its effectiveness as a clustering quality function for networks or graphs and because many fast, scalable heuristics already exist for maximizing network modularity (i.e. Louvain \cite{Blondel:2008}, Leiden \cite{Traag:2019}, etc.). We refer to this multi-view clustering technique as Multi-view Modularity Clustering (MVMC). The following figure, Figure \ref{fig:multi-view_clustering_overview}, displays the overall methodology.

\begin{figure}[ht]
    \centering
    \includegraphics[width=0.9\textwidth]{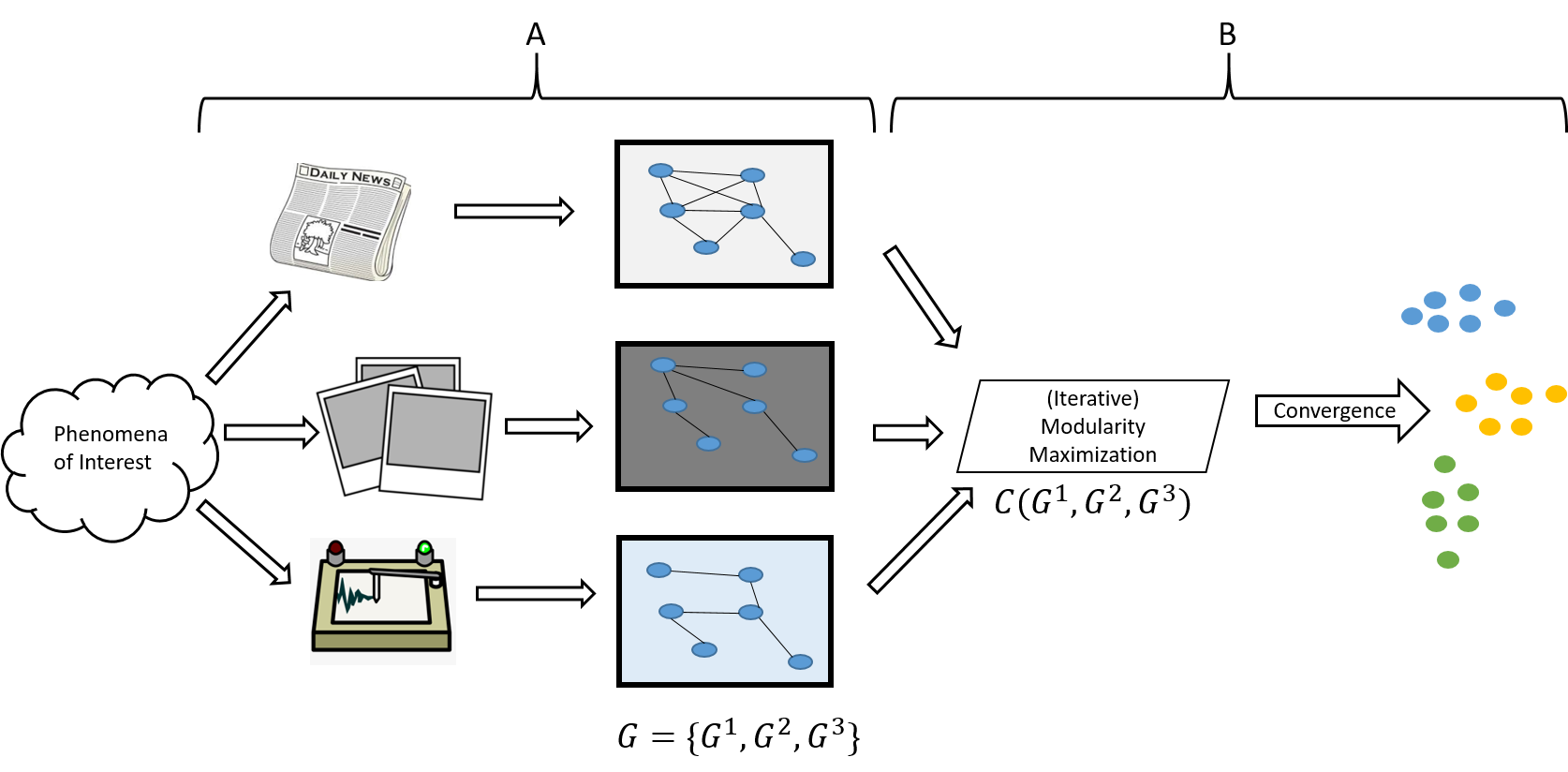}
    \caption[Overview of the Multi-view Modularity Clustering method]{Overview of the Multi-view Modularity Clustering (MVMC) method. the method works in two main steps. In step A, all of the views of the data are converted into graphs. In step B, these view graphs are collectively clustered using an iterative modularity maximization technique to produce clusters.}
    \label{fig:multi-view_clustering_overview}
\end{figure}

The methodology proposed in this work for clustering multi-view. social-based data consists of two steps. The first step is to form graphs for every view of the data (A in Figure \ref{fig:multi-view_clustering_overview}). This can done by the computationally quick heuristic of using a symmetric k-Nearest Neighbor graph, where $k$ is the square root of the number of vertices, $k=\lfloor \sqrt{n} \rfloor$, for modes that are not already networks or graphs \cite{Qiao:2018}. While this method can be used for creating the graphs, there are many other methods for inferring or learning graphs from data which could easily be used in this methodology \cite{Brugere:2018}, \cite{Qiao:2018}. Having formed graphs of all of the views, the next step is to use a modularity-based clustering procedure to simultaneously cluster all of the views (B in Figure \ref{fig:multi-view_clustering_overview}). This can be done by using a multiplex network modularity maximization procedure \cite{Mucha:2010}, \cite{Pamfil:2018}. 

\subsection*{Creation of the View Graphs}

The first step of the MVMC procedure is to create graphs of each of the views. For each view, a similarity graph was created. So, for each view graph, an edge represents how similar two objects are with respect to that view. For example, for the text view, an edge indicates that two hashtags share similar text in their tweets, or for the co-occurence view that two hashtags tend to occur with the same set of other hashtags. In order to measure similarity, I first transformed the view data from raw counts (i.e. the number of times a users uses a hashtag) to Term Frequency-Inverse Document Frequency (tf-idf) scores using,

\begin{equation}
w_{ij} = tf_{ij} \times \frac{n}{df_{j}}
\end{equation}

where $tf_{ij}$ is the number of terms (i.e. users, word tokens, URLs, etc.) occur with hashtag $i$, $df_{j}$ is the number of Hashtags that also feature the $jth$ term, and $n$ is number of hashtags. This transformation was done in order to down-weight those terms which are common across all hashtags, like `COVID19' and up-weight those terms which may be significant for cluster structure. While it has been noted in previous works that tf-idf can be insufficient for textual information for tweets, the primary reason for this insufficiency that individual tweets have very little text which can accompany them \cite{Vicient:2015}, \cite{Xiao:2014}. However, in this methodological set up, many tweets (often hundreds or thousands) are combined for each hashtag, so the text size limitation is not a significant issue. Furthermore, using tf-idf is also very scalable and does not require any kind of \textit{a priori} semantic knowledge database \cite{Vicient:2015}. So, for these reason, we have opted to use tf-idf as a means of processing the view data prior to learning the view graphs. Having applied tf-idf to all of the views, the similarity for each of the views was measured by cosine similarity:

\begin{equation}
s_{ij} = \frac{A_{i} . A_{j}}{||A_{i}|| \times ||A_{j}||}
\end{equation}

Since the data is large in the number of objects that need to be clustered (i.e. hashtags) --- on the order of tens of thousands --- the graph learning procedure needs to be a computationally efficient one \cite{Qiao:2018}. For that reason, we have adopted the heuristic procedure of creating a symmetric k-Nearest Neighbor Graph (k-NN) with the number of nearest neighbors as $k=\sqrt{n}$, where $n$ is the number of objects being clustered \cite{Maier:2011}, \cite{Maier:2009}. To symmetrize the k-NN the following step as used: once each object has been connected to $k$ of its nearest neighbors, we have adopted the average strategy, $A'=\frac{1}{2} A + A^{T}$, which is common in spectral clustering methods \cite{Qiao:2018}, \cite{Yu:2017}, \cite{Zhu:2014}. So, at the end of the graph learning step, there is a cosine-similarity weighted, undirected graph for each view of the data.

\subsection*{Multi-view Clustering}

In order to cluster the graphs of all of the views of the data, we use a modularity maximization technique. The function used in the optimization is a modified version of that proposed in \cite{Mucha:2010} for dealing with clustering multi-layer networks. The original multi-layer modualrity is given by:

\begin{equation}
\begin{split}
    Q &= \sum_{v=1}^m \sum_{ij \in E^v} [ A^v_{ij} - \frac{deg(i)^v \times deg(j)^v}{2 \sum E^v} ]\delta( {C^v_i, C^v_j}) + \sum_{v=1}^m \sum_{s\neq v} \sum_{i=1}^n \omega_i^{sv} \delta( {C^s_i, C^v_j}) \\
    Q &= Q_{intra} + Q_{inter}
\end{split}
\end{equation}

Where $A^v$ is the adjacency matrix for a view, $v$, $deg(i)^v$ is the degree of vertex $i$ in view $v$ (i.e. $deg(i)^v = \sum_{j=1}^n A^v_{ij}$), $E^{v}$ is the set of edges in the $v$th graph, $m$ is the number of views, $n$ is the number of vertices, the $\delta$ function is one if vertices $i$ and $j$ are in the same cluster, $\omega^{sv}$ controls the strength of vertices being in the same cluster between layers, and $C_i^v$ is the cluster that vertex $i$ belongs to in view $v$. So, this function can be broke into two parts: the first measures the modularity of of the clustering within each layer, $Q_{intra}$ and the second measures the the modularity of the clustering between layers, $Q_{inter}$. For multiplex clustering, or multi-view clustering, where clusters are not allowed to vary between views or layers, this modularity can be reduced to just the sum of the intra-layer modularities. Furthermore, it is also possible to weight the different views, or layers, differently as well. So, a weighted multi-view modularity would have the form of:

\begin{equation}
    Q = \sum_{v=1}^m w^{v} \sum_{ij \in E^v} [ A^v_{ij} - \frac{deg(i)^v \times deg(j)^v}{2 \sum E^v} ]\delta( {C_i, C_j})
\end{equation}

where $w^{v}$ is now the weight of a particular view. Despite its empirical successes and strong grounding in statistical physics, modularity does have an important shortcoming: modularity has a \textit{resolution} limit. In brief, the resolution limit is when modularity cannot detect clusters present in a network when the network is sufficiently large and communities are sufficiently small \cite{Lancichinetti:2011}, \cite{Fortunato:2007}. In order to address this shortcoming, several authors have proposed various means of correcting for different resolutions when clustering networks \cite{Traag:2011}, \cite{Santiago:2020}, \cite{Reichardt:2006}, \cite{Traag:2013}. One means of addressing the resolution limit is to add a parameter to the modularity to directly account for the resolution that may be present in a network:

\begin{equation}
    Q = \sum_{v=1}^m w^{v} \sum_{ij \in E^v} [ A^v_{ij} - \gamma^v \frac{deg(i)^v \times deg(j)^v}{2 \sum E^v} ]\delta( {C_i, C_j})
\end{equation}

where $\gamma^v$ is now a resolution parameter for each view graph that can be set depending on the number of clusters that are present in the network relative to the network's size \cite{Reichardt:2006}. This final equation is now the measure of cluster structure present in the multi-view data, and the objective function for modularity optimization procedures, like Leiden or Louvain.

One issue this final function of multi-view network modularity parameters raises is how to set the view weights and view modularities, without \textit{a priori} knowing what they should be. Previous works have found what the resolution parameters should be for a network given a clustering of that network. In order to set the the resolution parameter to an optimal value for a particular graph, the following function is used:

\begin{equation}
    \gamma = \frac{\theta_{in} - \theta_{out}}{\text{log} \theta_{in} - \text{log} \theta_{out}}
\end{equation}

where $\gamma$ is the resolution parameter, and $\theta_{in}$ and $\theta_{out}$ are the propensities of having edges internal to clusters or external to clusters respectively. This function was derived by relating modularity maximization to the planted partition Stochastic Block Model \cite{Newman:2016}, \cite{Pamfil:2018}. The intuition behind this function is that when there is a greater propensity to form edges internal rather than external to clusters that the resolution should be higher which would bias the function to find more tightly-knit and possibly smaller communities. Since this formulation relies on knowing the clusters to compute the $\theta$ values, it does not on first glance seem useful for actually clustering a graph. However, the function for computing the resolution has been used in an iterative fashion with modularity-based graph clustering to optimally cluster graphs in reasonably few (i.e. less than 20) iterations \cite{Newman:2016}, \cite{Pamfil:2018}. So, it is possible to iteratively cluster and then update the resolution parameter to both find the optimal clustering of the graph as well as its appropriate resolution parameter.

The same means of computing the optimal resolution parameter can be extended to compute the optimal view weights. Pamfil et al. used the same derivation process of finding the optimal resolution parameters from the relation of Reichardt and Bornholdt modularity to planted partition Stochastic Block Models to obtain the optimal weights for each of the layers in a multiplex graph as:

\begin{equation}
    w_v = \frac{\text{log} \theta^v_{in} - \text{log} \theta^v_{out}}{<\text{log} \theta^v_{in} - \text{log} \theta^v_{out}>_v}
\end{equation}

where $w_v$ is the weight given to a view, $v$, and $\theta^v_{in}$ $\theta^v_{out}$ are the propensities for edges to form internal to a cluster or external for the $v$th view, respectively. $<.>_v$ is the average across all of the views. The intuition behind this derivation is that those views with higher propensity to have edges internal to clusters versus having edges external to clusters relative to the average across all views will have higher weights. So, a view with a better than average propensity to have edges internal to clusters should be weighted more heavily in the modularity calculation for clustering. Once again, as with the resolution parameter, the weight parameter can be determined in an iterative fashion \cite{Pamfil:2018}.

Having defined the new objective functions, a new algorithm can then be developed to cluster multi-view data by a modularity maximization optimization. At a high level, the algorithm runs by first assigning staring resolution and weight parameters for every view (typically one) and then clustering the graph using a modularity maximization technique like Louvain \cite{Blondel:2008} or Leiden \cite{Traag:2019}. These clusters are then used to compute new resolution and weight parameters. This process is repeated until the resolution and weight parameters no longer change. In the event that the resolution and weight parameters do not converge (which can happen in practice \cite{Pamfil:2018}), the clustering with the highest modularity value is chosen as the final clustering. The following psuedocode, Algorithm \ref{alg:Multi-view_Modularity_Clustering} describes the algorithm in detail.

\begin{algorithm}
\caption{Multi-view Modularity Clustering (MVMC)}
\label{alg:Multi-view_Modularity_Clustering}
\begin{algorithmic}
\BState \textbf{input}:
\begin{itemize}
    \item Adjacency for each view: $A^{v}$
    \item Max number of iterations: $max\_iter=20$
    \item Starting resolutions: $\gamma_1^{v}=1$, $\forall v \in m$
    \item Starting weights: $w_1^{v}=1$, $\forall v \in m$ 
    \item Convergence tolerance: $tol=0.01$
\end{itemize}
\BState \textbf{output}: Cluster assignments

\State $clustering^* \gets None$
\State $modularity^* \gets - \infty$
\For{$i=1:max\_iter$}
    \State $clustering_i \gets cluster(A, w_i, \gamma_i)$
    \State $modularity_i \gets RBmodularity(A, clustering_i, w_i, \gamma_i)$
    \State $\theta_{in}, \theta_{out} \gets calculate\_thetas(A, clustering_i)$
    \State $\gamma_{i+1}^{v} \gets \frac{\theta_{in}^{v} - \theta_{out}^{v}}{\text{log} \theta_{in}^{v} - \text{log} \theta_{out}^{v}}$, $\forall v \in m$ 
    \State $w_{i+1}^{v} \gets \frac{\text{log} \theta^v_{in} - \text{log} \theta^v_{out}}{<\text{log} \theta^v_{in} - \text{log} \theta^v_{out}>_v}$, $\forall v \in m$ 
    \If{$abs(\gamma_{i+1}-\gamma_i)<tol$ AND $abs(weights_{i+1}-weights_{i})<tol$}
        \State $clustering^* \gets clustering_i$
        \State $modularity^* \gets modularity_i$
        \State BREAK
    \EndIf
    \If{$iter>=max\_iter$}
        \State $best\_iteration \gets argmax(modularity)$
        \State $clustering^* \gets clustering[best\_iteration]$
        \State $modularity^* \gets modularity[best\_iteration]$
    \EndIf
\EndFor
\State \Return $clustering^*$
\end{algorithmic}
\end{algorithm}

The algorithm begins by initializing all the resolution parameters, $\gamma^v_1$, and weight parameters, $w^v_1$ to one (or whatever the user may specify). The algorithm then goes on to cluster the view graphs, $A^{v}$, by a modularity maximization technique (i.e. Louvain, Leiden), $cluster()$, with the current resolution and weight settings. The output of this is then used to determine the propensities for internal edge formation $\theta_{in}^v$, and external edge formation, $\theta_{out}^v$ for each view. These values are then used to update the resolution, $\gamma^v$, and weight parameters, $w^v$, for each of the views. If the new weight and resolution parameters are the same as the previous ones (within tolerance), the algorithm then exists and returns the final clustering. If the algorithm fails to converge to stable resolution and weight parameters, within the maximum  number of iterations allowed, then the algorithm returns whichever clustering produced the highest modularity. Note, that modularity for this algorithm is the view-weighted, Reichardt and Bornholdt modularity, which incorporates the view resolutions. 

One of the important elements in the aforementioned algorithm, Algorithm \ref{alg:Multi-view_Modularity_Clustering}, is the computation of the edge propensities, $\theta$. In order to calculate these edge propensities, we follow the guidance outlined in previous works and assume edges form by a degree-corrected model \cite{Newman:2016}, \cite{Pamfil:2018}. Given a degree corrected model, the expected number of edges that occur internal to clusters is given by:

\begin{equation}
\begin{split}
    e_{in} & = \frac{1}{2}\sum_{c}\sum_{ij \in E} \theta_{in} \frac{deg(i)deg(j)}{2 \sum E} \delta(C_i,C_c)\delta(C_j,C_c) \\
    & = \frac{\theta_{in}}{4\sum E} \sum_{c}\kappa_{c}^2
\end{split}
\end{equation}

where $c$ is a cluster and $\kappa_c = \sum_{i} deg(i)\delta(C_i,C_c)$, or the sum of the degree of the vertices within cluster $c$. Using the observed number of edges internal to the clusters for the expected number of edges internal to clusters, $e_{in}$, this equation can then be used to calculate the propensity to form edges internally as:

\begin{equation}
    \theta_{in} = \frac{e_{in}}{\sum_c \frac{\kappa_c^2}{4 \sum E}}
\end{equation}

Similar to the propensity to form edges internally, the propensity to form edges externally can be derived from the expected external edges under the degree-corrected model as:

\begin{equation}
    \theta_{out} = \frac{\sum E-e_{in}}{\sum E -\frac{\sum_c \kappa^2_c}{4 \sum E}}
\end{equation}

With these equations, and the assumption of edges forming by a degree-corrected model, the propensities for edges to occur internal or external to a cluster can be calculated. It is important to note, however, that these equations assume the \textit{observed} edges internal or external to the clusters are equal to the \textit{expected} edges internal or external to clusters. In practice, these values may actually differ. For example, if every vertex ends up in its own cluster, than no edges will be internal which will lead to the propensity internal term, $\theta_{in}$ to become zero, and the resolution and weight updates to fail. A similar problem can occur if all the vertices end up in one cluster and so no external edges occur. While at first glance these examples may appear to be edge cases, these problems can also occur in more important cases. For example, if the graph consists of a series of disconnected cliques. The optimal clusters in this situation would then be to put all of the cliques within their own, separate clusters. However, this would result in there being no external edges, and so the resolution and weight updates would fail. In order to address these shortcomings, we have chosen to have a small value substitute for the propensities if there are either no internal edges and/or no external edges. So, while the observed edges can act as a proxy for the expected edges in most cases, the expected edges will always be a nonzero number. With these corrections, the algorithm for computing the propensity values is given in pseudocode by Algorithm \ref{alg:calculation_of_propensities}.

\begin{algorithm}
\caption{Calculation of Edge Propensities}
\label{alg:calculation_of_propensities}
\begin{algorithmic}
\BState \textbf{input}:
\begin{itemize}
    \item Adjacency for each view: $A^{v}$
    \item clustering: $C$
\end{itemize}
\BState \textbf{output}: Internal and external edge propensities ($\theta_{in}, \theta_out$)
\For{v = 1:m}
    \State $e_{in} = 0$
    \State $\kappa^2 =[]$
    \For{c = 1:|C|}
        \State $e_c = \sum E^{v}_c$
        \State $e_{in} += e_c$
        \State $\kappa^2.append((\sum_{i \in V^{v}_c} deg(i))^2)$
    \EndFor
    \If{$e_{in} =0$}
        \State $\theta^{v}_{in} \gets \frac{1}{|E^{v}|}$
    \Else
        \State $\theta^{v}_{in} \gets \frac{e_{in}}{\sum \frac{\kappa^2}{4 \sum E^{v}}}$
    \EndIf
    \If{$e_{in} ==\sum E^{v}$}
        \State $\theta^{v}_{out} \gets \frac{1}{|E^{v}|}$
    \Else
        \State $\theta^{v}_{out} \gets \frac{\sum E^{v}-e_{in}}{\sum E^{v} -\sum \frac{\kappa^2}{4 \sum E^{v}}}$
    \EndIf
\EndFor
\State \Return $\theta_{in},\theta_{out}$
\end{algorithmic}
\end{algorithm}

The algorithm goes through each graph to calculate the propensities for each graph separately. For each graph, the algorithm begins by calculating the number of internal edges and the degree-corrected, null-model terms (i.e. $\kappa^2$) for each of the clusters. Then, the algorithm checks as to whether the graph is directed or undirected and whether there are no internal or external edges and then calculates the final propensities for that view graph, $\theta_{in}^v, \theta_{out}^v$. Once the propensities have been calculated for all of the view graphs, these are then returned. With the calculation of the edge propensities, $\theta$, we now have everything needed to perform multi-view clustering of the hashtags based upon the four different views of their usage.

For clustering the COVID-19 hashtag data, the following parameters for MVMC were used: The initial weights and resolutions were all set to one. The convergence tolerance for the resolutions was set to 0.3 and for the weights to 0.1. And, the procedure was allowed to run for a maximum of 20 iterations.

\section*{Results}

In this section, we describe the multi-view clustering results on the COVID-19 twitter data. In the first section, we provide an overview of the clustering results. In the second section, we detail the results of learning graphs on the different views of the data and the insights the graphs give about the data and the cluster structures present within the data. In the third section, we analyze the temporal patterns within the hashtag cluster data. In the fourth section we analyze the user bases of the different clusters of hashtags. Finally, in the fifth section we do an in depth analysis of some of the interesting clusters identified within the data.

\subsection*{Overview of Multi-view Clustering Results}

Multi-view clustering of the COVID-19 twitter data extracted between 20 and 160 clusters of hashtags per day, with a varying size of the clusters. These clusters varied in size and also in number between different days. The following figure, Figure \ref{fig:cluster_stats}, displays plots of the daily number of clusters and daily cluster size statistics over the full 90 day period.

\begin{figure}[ht]
    \centering
    \includegraphics[width=0.9\textwidth]{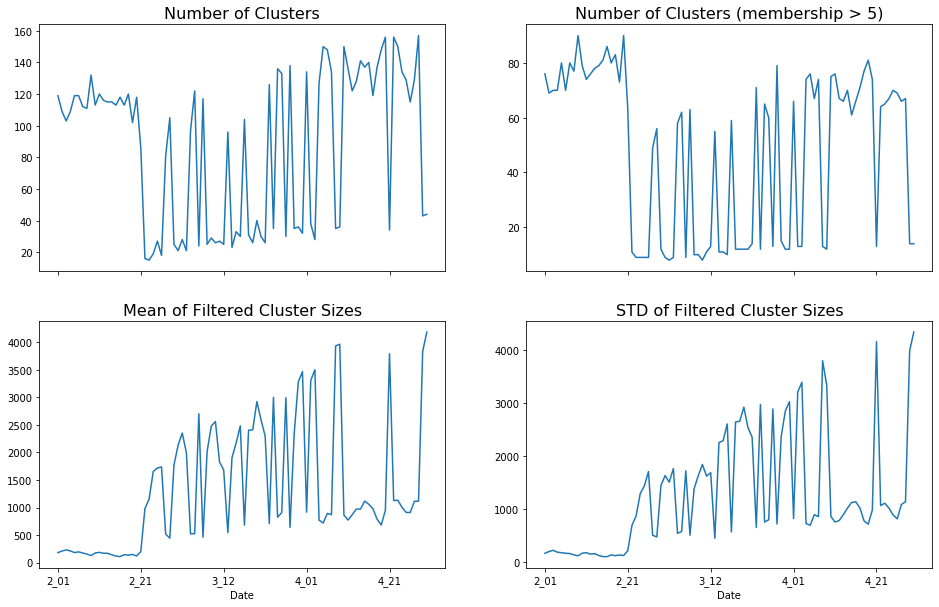}
    \caption[Daily clustering statistics on clusters produced by the MVMC technique on four views of the hashtags]{Daily clustering statistics on clusters produced by the MVMC technique on four views of the hashtags. The daily clusters display three different, temporal patterns of clustering.}
    \label{fig:cluster_stats}
\end{figure}

There is a presence of several small clusters within the daily clusterings. First, on any given day there were between 20 and 80 clusters that had a size of less than five objects. These clusters were almost exclusively composed of either small, locality-specific hashtags, non-English language hashtags, or hashtags that seem to have little relevance to the pandemic. For example, a particular local business, like a car dealership may have a hashtag that is used in a tweet that happens to mention one of the key terms, which was then retweeted by local residents, which would cause it to be above the initial screening criteria of being used in more than three tweets, but otherwise bear little relation to how any of the other hashtags in the data are used. These small clusters are also often composed of objects that are isolates within one or more of the view graphs, which is what makes them difficult to group into larger clusters. This result illustrates an important point about multi-view clustering of real-world data; the data is often messy and incomplete and requires some degree of additional processing. These arbitrary clusters do not actually contribute much to understanding the data, or the macro-cluster structure of the data, or use of hashtags beyond recognizing that the collection process can produce some noise in clustering results. Removing these small clusters does not affect the overall patterns existing within the clusters, and makes interpretation of the clusters easier. 

Second, there are dynamic patterns within the clusters. The clusters start off as many in number and small in size and become fewer in number and larger in size as time passes. Up until the end of February, there are around 80 non-arbitrary clusters with a size of around 250 hashtags. This pattern changes at the end of February where the number of daily hashtag clusters decreases but the size of these clusters increases. This change in the clustering structure over time may indicate that the use of hashtags and their associated discussions begin to congeal into larger discussions over the course of the pandemic. Additionally, there can be large oscillations in numbers of clusters and sizes of clusters between any given set of days. While, there is an increasing trend toward fewer and larger clusters, there are oscillations present within the data, especially during the middle of the time period, around the month of March. While it is not quite clear why these oscillations occur, it was noted in the description of the data that there are weekly periodic patterns within the number of unique hashtags used over time. So, its possible these oscillations are in part due to cyclical, time-dependent patterns in twitter use. This dynamic nature in the clusterings will be further investigated in an upcoming section.

\subsection*{Graph Learning Results}

In order to better understand the clustering results, we now turn to the analyzing the view graphs. As was described in the methodology section, the graphs for all of the different views were created by a heuristic symmetric k-NN graph learning procedure. This procedure is meant to learn a graph that represents the data. So, graph-theoretic and network science measures can be used to analyze the graphs and thereby better understand the data. The following figures, Figure \ref{fig:graph_stats}, display some important graph properties of the different view graphs for each of the daily data sets.

\begin{figure}[ht]
    \centering
    \includegraphics[width=0.9\textwidth]{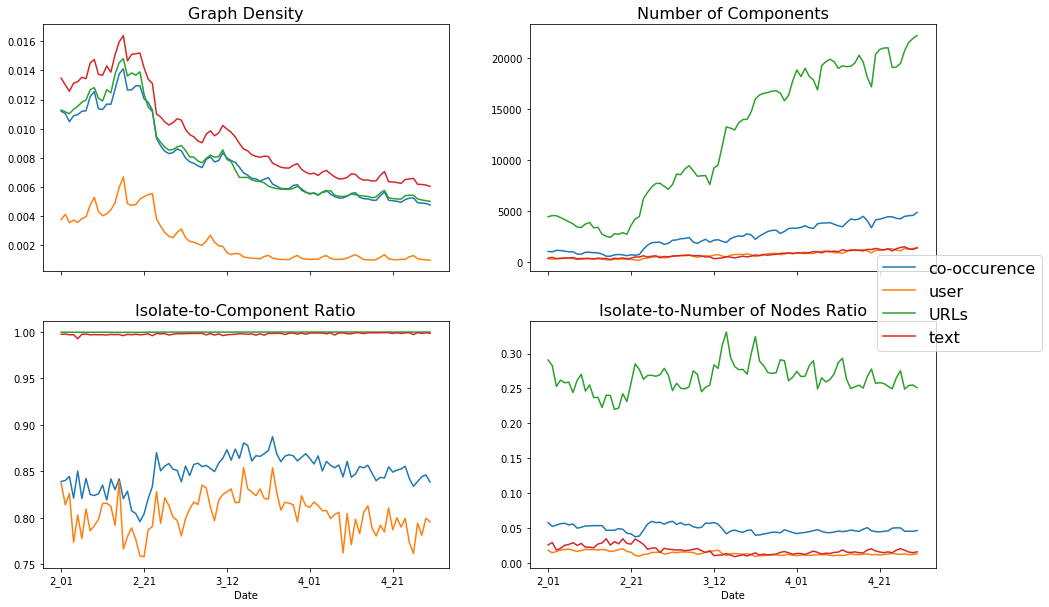}
    \caption[Graph metrics for each of the daily view graphs]{Graph metrics for each of the daily view graphs. All views but the URL view display useful graph properties for clustering the data.}
    \label{fig:graph_stats}
\end{figure}

From the graphs, one can first observe that the graph densities follow a pattern that would be expected from the the number unique of hashtags. Density initially increases slightly, and then decreases as time moves forward. From the section on the data it is also easy to observe that this pattern is roughly the inverse pattern of the number of daily unique hashtags. So, as expected from observations of real-world networks, as the number of nodes --- or unique hashtags, in this case --- increases, the density decreases \cite{Newman:2010}. Additionally, the text view is consistently the most dense graph while the shared users graph is the least dense graph. Since these graphs were created by a symmetric k-NN graph learning procedure, most of difference in density for the users view is from more groups of overlaps, around the size of $k$, in users between hashtags. When there is more groups of overlaps around the size of $k$ on the features, nearby objects in a k-NN  graph tend to be within the $k$ nearest neighbors of each other and that there are only about $k$ neighbors for any given point, which results in fewer edges forming overall once the graph is symmetrized. Conversely, when there are more similar neighbors for each object than the value of $k$, there will be more edges in the symmetrized k-NN graph as two nearby objects, while very similar to each other, may not be within the top-$k$ nearest neighbors of each other.

Second, the component statistics vary considerably between the different views of the data. The URLs view, which measures similarity between hashtags if they co-occur in tweets with the same URLs, has far more components than the other views, and these components --- with the exception of one major connected component --- are almost always isolates. This is to say that of the hashtags that co-occur with a URL, there are a fair number of URLs that only co-occur with a particular hashtag. This result is also partly an artifact of resolving the Twitter shortened URLs; some of the shortened URLs were unable to resolve to the non-Twitter URL, and Twitter does not always have the same shortened URL for any given URL. Thus, we would expect the URL view to not be particularly useful in finding communities of hashtag use. While much lower on the number of components, a similar pattern is observed with the text mode; those hashtags which are not part of the major connected component are almost always isolates. These hashtags are often rarely used hashtags, typically because they are a common misspelling of a popular hashtag or are a less popular hashtag that occurs with non-English text or no text (i.e. just the hashtag by itself was tweeted). Also, it should be noted that the co-occurrence view, which measures similarity between hashtags based on the other hashtags that those hashtags appear with in a tweet with, has a slightly higher number of components and percentage of isolates than either the user or text views. This is due to the fact that some hashtags never co-occur with another hashtag in a tweet. finally, it is worth noting that for all of the views, the number of components increase with time. These observations suggest that the use of hashtags is becoming more divided into distinct and non-interacting communities.

\subsection*{Temporal Ensemble Clustering Results}

As has been noted throughout the results section and the data section, the usage of hashtags seems to have some temporal trends and changes over the period of investigation. So, in this section we will analyze the daily clusterings produced by MVMC to understand the temporal nature of the hashtag clusters and hashtag usage. To assess any possible temporal patterns that could exist within the daily clusterings, we analyzed how similar the daily clusterings are to each other. Comparing the similarities between the clusterings can give insight into how stable both the usage of hashtags and the broader discussion topics which use the hashtags are between days. In order to compare the daily clusterings, we opted to use the Adjusted Rand Index (ARI) \cite{Hubert:1985}, which provides a value between zero and one that expresses how similar two clusterings are. In order to measure the ARI between all of the daily clusterings, each clustering has to have the same objects, or hashtags, as every other clustering. So, a set of all of the hashtags used across the entire data set was collected from the filtered hashtag clusters (the filtered hashtag clusters are those clusters which have at least five hashtags in them, for each day). For each daily clustering, if a particular hashtag was not present on that day, it was added to the daily clustering and assigned a dummy cluster label. So, for each day, the clusterings have the same hashtags and those hashtags which do not occur on a particular day are all assigned the same dummy label for that day. Having cross-leveled the hashtags across all of the days in the data set, the pairwise ARI between each day's clustering and every other day's clustering can then be computed. The pairwise ARIs between the daily clusterings are summarized in the following figure, Figure \ref{fig:clustering_heatmap}. 

\pagebreak
\begin{figure}[ht]
    \centering
    \includegraphics[width=0.9\textwidth]{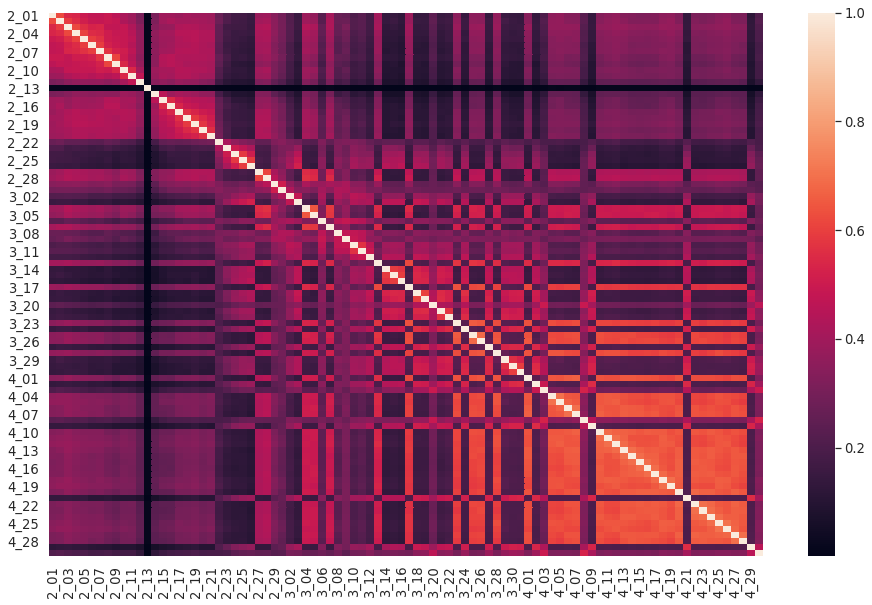}
    \caption[Heat map of the ARI values between every daily clustering]{Heat map of the ARI values between every daily clustering produced by MVMC with every other daily clustering. The heat map shows there are some regions where the consecutive days are more similar to each other than to other days. Examples include early to mid-February in the top left and mid to late-April in the bottom right.}
    \label{fig:clustering_heatmap}
\end{figure}

\begin{figure}[ht]
    \centering
    \includegraphics[width=0.9\textwidth]{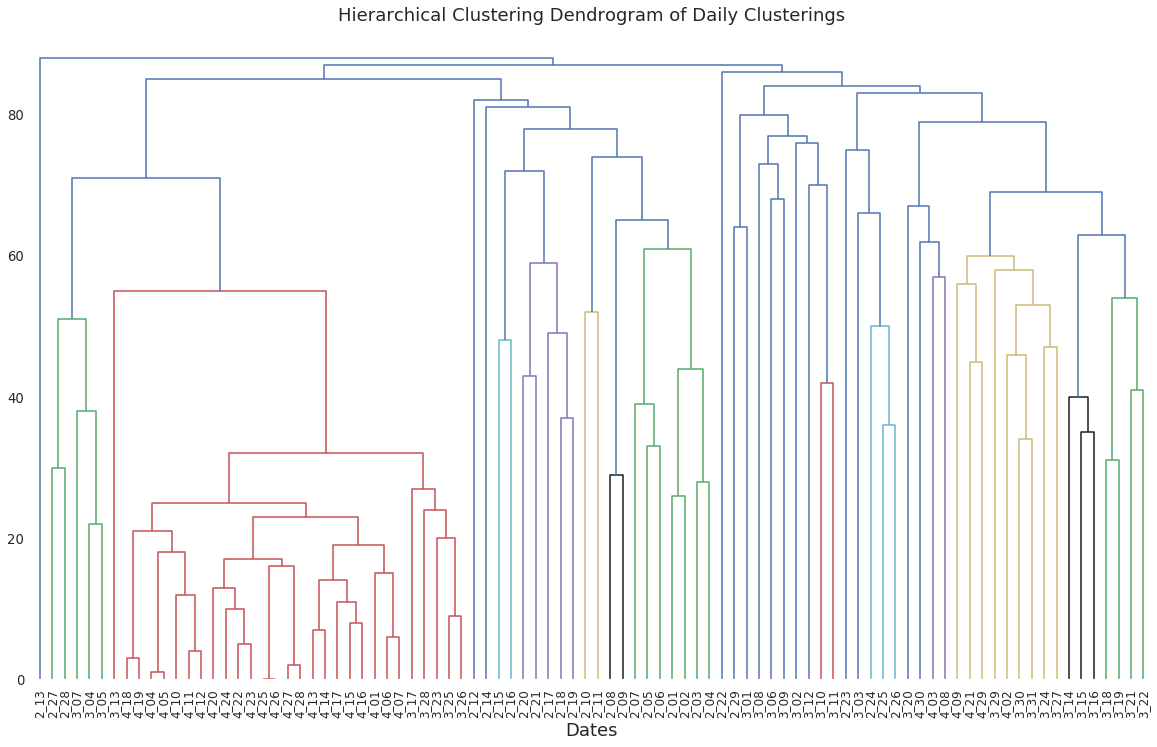}
    \caption[Cluster Dendrogram of the daily clusterings]{Cluster Dendrogram of the daily clusterings produced by Agglomerative Hierarchical Clustering with ARI as the measure of similarity. The Dendrogram shows 3 main clusters with some outlier daily clusterings.}
    \label{fig:clustering_dendrogram}
\end{figure}

From the figure it can be observed that there are some block structures present within the data along with some outlier clusterings. For example, early to mid-February has clusterings which, with the exception of the 13 of February, are all more similar to each other than to any other days' clusterings. This temporal pattern was similarly observed in the clustering overview statistics and MVMC performance statistics. Additionally, clusterings in April also tend to display a block structure whereby the clusterings are more similar to each other than to any other days' clusterings. Outside of these block structures, there are also some outlier clusterings that are not more similar to those clustering which are temporally close. The 13th of February provides an extreme example in that it has very low similarity to every other clustering. Since it appears there are clusters of daily clusterings present within the data along with a pairwise measure of similarity, the clusterings can themselves be clustered. To cluster the daily clusterings, we used the pairwise ARI scores in Agglomerative Hierarchical Clustering with average linkage. It should be noted that clustering a set of clusterings has been used to analyze other temporal streams of data in order to understand dynamic trends within the data \cite{Magelinski:2019}, \cite{Masuda:2019}. The following the figure, Figure \ref{fig:clustering_dendrogram}, displays the full dendrogram for the clustering of the daily clusterings. 

\pagebreak

From the dendrogram it can be observed that there are indeed clusters of the daily clusterings. And, these clusters tend to consist of temporally nearby clusterings. Thus, it would seem that there are temporal meta-clusters of daily clusterings present in the data. To analyze these temporal meta-clusters, we first clustered the clusterings. Based the dendrogram, the clusterings were divided into 5 clusters. Note that the division of the daily clusterings is done without respect to time, but rather is done only on the pairwise ARI between the daily clusterings. Having partitioned the daily clusterings into meta-clusters, we can then see if these meta-clusters correspond to any time periods within the data. The following figure, Figure \ref{fig:clustering_clusters_timeline}, displays a plot of the daily clusterings over time versus the meta-cluster that the daily clusterings were partitioned into.

\begin{figure}[ht]
    \centering
    \includegraphics[width=0.9\textwidth]{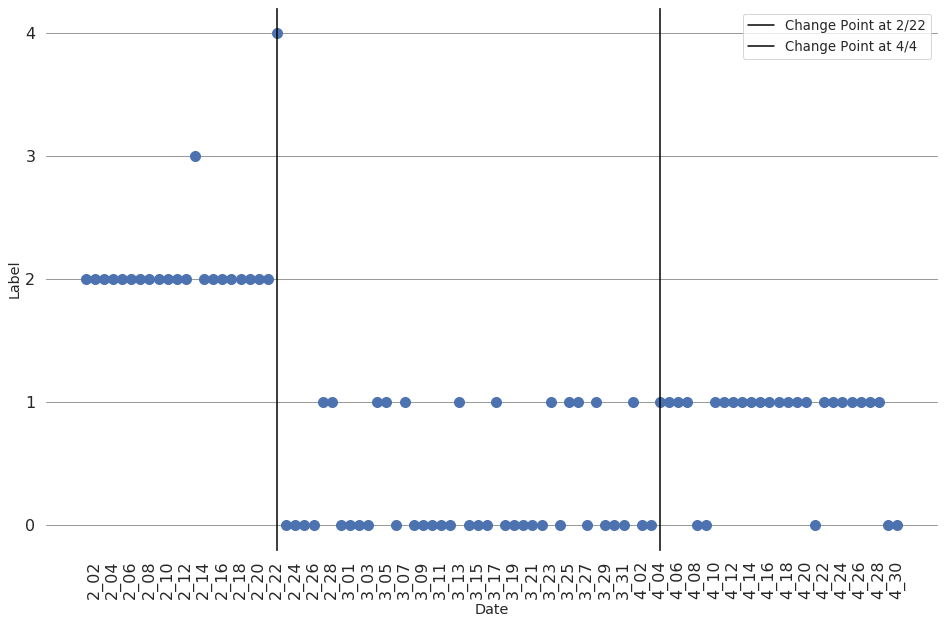}
    \caption[Plot of daily clustering versus what meta-cluster that daily clustering falls into]{Plot of daily clustering versus what meta-cluster that daily clustering falls into. Generally, the meta-clusters of the daily clusterings have distinct temporal bounds indicating that daily clusterings have a macro, temporal structure to them.}
    \label{fig:clustering_clusters_timeline}
\end{figure}

From the figure, there is an observable temporal pattern to the meta-clusters. There is a meta-cluster (label 2) that is exclusively composed of the clusterings from early to late-February. The other two meta-clusters largely contain clusterings from either February 23rd to April 3rd, or April 4th to April 30th. There are also two outlier meta-clusters that consist of only one date, February 13th and February 22nd (labels 3 and 4, respectively). Thus, there is a macro temporal pattern within the daily hashtag clusterings. This temporal pattern in the hashtag clusterings reflects a similar pattern regarding user hashtag ratios that were observed in the data section. In fact the middle period of clusterings, which is the least distinct of the 3 time periods --- having meta-cluster 0 and 1 members --- corresponds closely to the time period where there were many oscillations in MVMC performance. So, not only does hashtags usage on the user level change over the course of the pandemic, but also the topical groups of hashtags also change over the course of the pandemic. And, it would seem there are two stable periods of hashtag usage that occur during the early stages of the pandemic in February and after the pandemic had been raging globally for some time, in early April.

\subsection*{Analysis of Temporally-Ensembled Clusters}

Having observed three distinct time periods of hashtag clusters, we would now like to get a better sense of how these periods differ in terms of hashtag usage. In order to better understand the hashtag clusters from the different time periods, each of the clusterings making up a meta-clustering are transformed into one clustering through cluster ensembling. This is done for two reasons: First, it makes the selection of which days and which clusters to analyze less arbitrary by reducing the number of clusters that need to be analyzed. Second, producing an ensemble clustering for each of the time periods can better mitigate any daily idiosyncrasies that could affect any given clustering on any given day, and thereby produce a better overall clustering that represents the whole time period. To produce ensemble clusters for each time period the BGPA technique, which clusters the object-by-cluster graph, was used \cite{Fern:2004}. This technique was used as it is a very scalable cluster ensembling technique in terms of the number of objects (i.e. hashtags) which it can handle \cite{Boongoen:2018}. The following table, Table \ref{tab:cluster_ensembling_results}, displays a summary of the results for each of the ensembled period meta-clusters.

\pagebreak
\begin{longtable}{|lllll|}
\hline
Period                                         & \begin{tabular}[c]{@{}l@{}}Number \\ of Clusters\end{tabular} & \begin{tabular}[c]{@{}l@{}}Average\\ Size of\\ Clusters\end{tabular} & \begin{tabular}[c]{@{}l@{}}STD Size\\ of Clusters\end{tabular} & \begin{tabular}[c]{@{}l@{}}Average Internal\\ ARI\end{tabular} \\ \hline
\multicolumn{1}{|l|}{February 1 - February 22} & 14                                                            & 751                                                                  & 1104                                                           & 0.416                                                          \\
\multicolumn{1}{|l|}{February 23 - April 3}    & 13                                                            & 747                                                                  & 1093                                                           & 0.348                                                          \\
\multicolumn{1}{|l|}{April 4 - April 30}       & 16                                                            & 566                                                                  & 691                                                            & 0.536                                                          \\ \hline
\caption[Cluster statistics of the ensembled clusterings of the daily clusterings for each time period]{Cluster statistics of the ensembled clusterings of the daily clusterings for each time period. The ensembled clusterings display similar temporal trends to the other analyses of the data in that attributes like the number of clusters present in each period roughly reflect the numbers of daily clusterings and unique hashtag usage during these periods.}
\label{tab:cluster_ensembling_results}
\end{longtable}

The ensembled, period meta-clusters produced somewhat different cluster structures for the different time periods. The third period had slightly more clusters and smaller and more regularly sized clusters than the other time periods. This time period also had a a higher clustering similarity, in terms of the average pairwise ARI between its constituent daily clusterings. For the first and second time periods, there was often one large cluster that had a size of around 4,400 hashtags while the largest cluster in the third time period was 3,025 hashtags. So, the third time period has a more balanced cluster structure, across the entire time period, than the other two time periods. 

To get a better idea of the ensembled clusters, the following table, Table \ref{tab:qualitative_meta_clusters_assessment} provides a qualitative assessment of the topic of cluster as well as how focused and easily assignable a topic is to each of the clusters in each of the time periods. 

{\small
\begin{longtable}{|lll|}
\hline
Period 1 & General Topic                                                          & \begin{tabular}[c]{@{}l@{}}Focus Level\\ of Cluster\end{tabular} \\ \hline
0        & Multi-language, general use hashtags                                   & low                                                              \\
1        & Multi-lingual coronavirus-specific hashtags                            & low                                                              \\
2        & News resources related hashtags                                        & medium                                                           \\
3        & Chinese focused hashtags (often of negative sentiment)                 & high                                                             \\
4        & Thai related hashtags                                                  & medium                                                           \\
5        & Economy/Commerce related hashtags                                      & high                                                             \\
6        & U.S. Politics related hashtags                                         & high                                                             \\
7        & technology and business related hashtags                               & high                                                             \\
8        & Asian-languages hashtags                                               & high                                                             \\
9        & Multi-lingual, anti-racism and health news related hashtag             & medium                                                           \\
10       & French language and European related hashtags                          & medium                                                           \\
11       & Italian language hashtags                                              & high                                                             \\
12       & Arabic script hashtags                                                 & high                                                             \\
13       & All-caps hashtags with some Syrian Civil War hashtags                  & medium                                                           \\ \hline
Period 2 & General Topic                                                          & \begin{tabular}[c]{@{}l@{}}Focus Level\\ of Cluster\end{tabular} \\ \hline
0        & News and U.S. Politics related hashtags                                & medium                                                           \\
1        & News-resources related hashtags                                        & low                                                              \\
2        & Commerce, Economy, and technology related hashtags                     & medium                                                           \\
3        & Asian-languages hashtags                                               & medium                                                           \\
4        & Spanish language hashtags                                              & medium                                                           \\
5        & French language hashtags                                               & high                                                             \\
6        & Italian language hashtags                                              & high                                                             \\
7        & Turkish language with some conspiracy theory related hashtags          & medium                                                           \\
8        & Online entertianment related hashtags                                  & hmedium                                                          \\
9        & German language hashtags                                               & medium                                                           \\
10       & Arabic script  and   middle-east related hashtags                      & high                                                             \\
11       & Australian and British news related hashtags                           & high                                                             \\
12       & Online education related hashtags                                      & high                                                             \\ \hline
Period 3 & General Topic                                                          & \begin{tabular}[c]{@{}l@{}}Focus Level\\ of Cluster\end{tabular} \\ \hline
0        & Spanish language and many general hashtags                             & low                                                              \\
1        & Multi-lingual, coronavirus and general hashtags                        & low                                                              \\
2        & Commerce, Economy, and technology related hashtags                     & medium                                                           \\
3        & Multi-lingual, coronavirus and general hashtags                        & low                                                              \\
4        & British news, anti-racism, and medical science related   and  hashtags & medium                                                           \\
5        & U.S. Politics related hashtags                                         & high                                                             \\
6        & Arabic-script and location related hashtags                            & medium                                                           \\
7        & Chinese focused hashtags (often of negative sentiment)                 & high                                                             \\
8        & French language hashtags                                               & high                                                             \\
9        & Italian language hashtags                                              & high                                                             \\
10       & Canadian and climate change related hashtags                           & high                                                             \\
11       & German language hashtags                                               & medium                                                           \\
12       & Asian languages hashtags                                               & high                                                             \\
13       & Indonesian and surrounding countries related hashtags                   & medium                                                           \\
14       & Thai language hashtags                                                 & high                                                             \\
15       & Turkish language hashtags                                              & high                                                             \\ \hline
\caption[Topical labels for the temporally-ensembled meta-clusters]{Topical labels for the temporally-ensembled meta-clusters. Some clusters had much more focused and readily defined topics than did others. Also, some topics are persistent throughout all three time periods, while some only exist in a time period.}
\label{tab:qualitative_meta_clusters_assessment}
\end{longtable}
}

Looking at the hashtags present in the clusters of the different ensembled meta-clusterings revealed both persistent topical groups and ones which change over time. In every time period, there is always a cluster that has hashtags for breaking news or news sources, a cluster that has business and commerce related hashtags, a cluster that has U.S. Politics-related hashtags, and foreign language clusters --- most notably Italian, German, and Spanish. These topical groups indicate that conversations about the global economy, news, and U.S. politics have remained important and consistent topics throughout the pandemic, and that even with English-language collection terms, the discussions occurring around the COVID-19 pandemic are international in nature. In addition to these persistent clusters there are also transient cluster topics that emerge in some time periods but not others. For example, in time period two there is a cluster of hashtags dedicated to online education and a cluster of hashtags concerning online entertainment and entertainment services (i.e. Hulu, Netflix). Both first and third periods contain clusters with negative sentiment hashtags toward the Chinese government and in support of Hong Kong protests. Overall, there are consistent topical clusters of discussion and other topics which emerge and disappear over time.

\subsection*{User-base Analysis of Temporally Ensembled Clusters}

In order to get a better sense of the hashtag clusters found through multi-view clustering and temporal ensembling, we can analyze the users that use the hashtags. In particular, it is of interest to observe whether those individuals which most use a hashtag also frequently use other hashtags from the same cluster. Presence of a small number of users being most active in the use of the hashtags could give insight into whether the topical conversation is being driven by a small group of users or is more of an open, less centrally-dominated topical discussion. To do so, we first found the top third of users for each hashtag in each cluster, across all periods, which we refer to as the `top users.' We then analyzed the number of unique top users for each cluster in each time period. The number of unique top users within any given topical hashtag group can give insight into whether there is a diverse user-base driving the topical discussion or not. The following figure, Figure \ref{fig:num_top_users}, shows the number of unique top users for each cluster in each time period.

\begin{figure}[ht]
    \centering
    \includegraphics[width=0.9\textwidth]{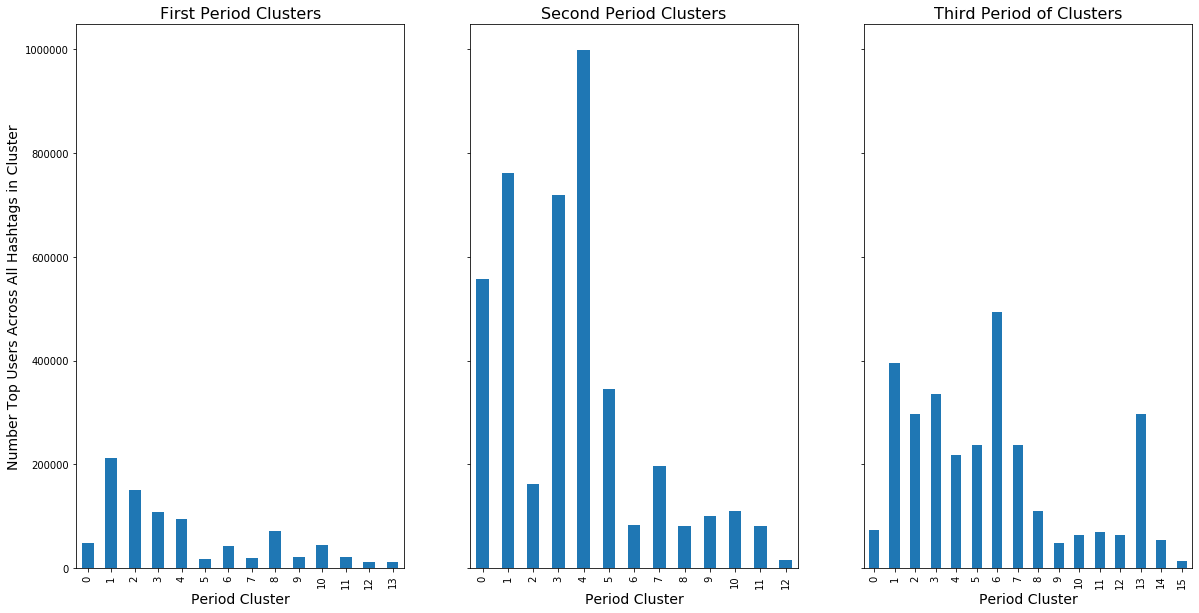}
    \caption[Number of unique users for each cluster of hashtags in each time period]{Number of unique users for each cluster of hashtags in each time period. There are distinct differences between the number of unique users between and within periods. These differences are largely not driven by the number of hashtags present within a cluster, but rather with the topic of those hashtags.}
    \label{fig:num_top_users}
\end{figure}

From the figure it can be observed that there are differences in user bases both between time periods and between clusters. The first and third time periods have generally fewer unique users in each of their clusters than the second time period. This due in a large part to the previous observation that there are more unique users in general on any given day during the second period than there are for the first or third periods. It is worth noting, however, that the total ratio of unique users to total users in both the second and third periods are about the same at $0.202$ and $0.206$ respectively. That is to say that even as the number of unique users decreases slightly in the third period and that the clusters in the third period do not have as many unique users, the period retains a relatively high number of unique users across the time period. Additionally, there are distinct differences in the number of unique top users between clusters within any of the time periods. This is especially true for the second time period. Generally, this difference in unique users is only partly accounted for by a difference in the size of the clusters as the 0th and 1st clusters are always the largest in any given period but do not have the greatest number of unique users for those time periods. This discrepancy in numbers of unique top users is also a result of the generality of the particular topic of the clusters, with those clusters having more general topics having higher numbers of unique top users. For example, the 4th cluster in the second period contains many hashtags, from many languages, which describe COVID-19, such as `$\#$COVID-19', `$\#$covid19', or `$\#$COVID---19.' So, the topical clusters of hashtags have differences in their top users with some clusters having a very small top user base and others, a larger one.

One of the issues with just looking at the the number of unique top users is that hashtags have different numbers of users in general. So, a hashtag could be completely used by a different user in each use, but the hashtag itself is not widely used, which would result in that hashtag having a small unique top user base. This effect extends to clusters where there are clusters of generally less used hashtags. So, I also use a \textit{top user score} for each cluster which compares the ratio of the number of top unique users for the hashtags versus the number of top unique users if there was no overlap between the top unique users of the hashtags. As a mathematical expression, this top user score for a cluster is given by:

\begin{equation}
    \text{top user score}^{c} = \frac{\sum\bigcap_{i=1}^r top\_users_i^{c}}{\sum_{i=1}^{r}\sum top\_users_i^{c}}
\end{equation}

where $c$ is a particular cluster, $r$ is the number of hashtags present in cluster $c$, and $top\_users_i^{c}$ is the set of users for hashtag $i$ in cluster $c$. The following figure, Figure \ref{fig:top_user_overlap_scores}, displays the top user scores for each of the clusters in each of the time periods.

\begin{figure}[ht]
    \centering
    \includegraphics[width=0.9\textwidth]{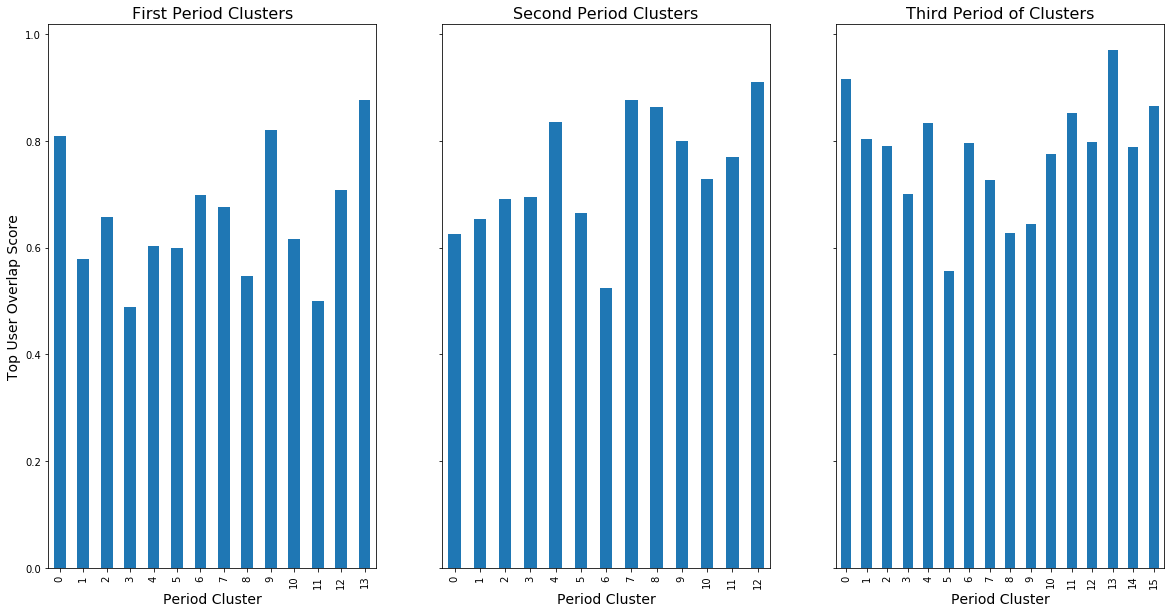}
    \caption[Top user scores for each cluster]{Top user scores, which compares the number of unique users for each of clusters to what would be expected if there were no overlap in the unique users between hashtags from the same cluster. Using this normalized measures allows us to observer differences in diversity in user bases of clusters which may have a small, but diverse, number of users}
    \label{fig:top_user_overlap_scores}
\end{figure}

From the figure, normalizing the unique top users by the actual hashtag usage across the clusters produces some different results than the previous figures. Some of the smaller clusters with a small user base can actually have a very diverse user base. For example, period one cluster 13, which focuses around the Syrian Civil War, has the highest top user score but a relatively small number of unique users. Another example is period 2 cluster 12, which focuses exclusively on online education-related hashtags, has a small number of unique users but the most diverse user base for the second time period. So, some clusters which can be small in the number of users can have very different users using the hashtags. This observation would suggest that only analyzing hashtags by creating hashtag-to-hashtag networks based on users is actually insufficient to find clusters of hashtags, which was a similar result observed by other authors \cite{Vicient:2015}. Additionally, some of the more mid-sized clusters can have less diverse user bases. For example, period one cluster 3, which focuses on hashtags critical of the Chinese government, has the least diverse user base in the first period, but a fair number of unique users. Period two cluster 6, which has many Italian-language hashtags, and period 3 cluster 5, which focuses on U.S. politics, have similar patterns. So, the topic of of a cluster tends to drive how diverse the user base of that cluster is, and not the number of hashtags or even the number of unique users.

To further explore the nature of the unique users of hashtags within clusters, we can look at how unique each hashtag's user base is within each cluster. In order to better understand the user base of a particular hashtag we calculated the ratio of the number of unique users that use a hashtag versus the number of times a hashtag is used in a given period. These hashtag user scores can then be combined to analyze each cluster within each of the time periods. The following figure, Figure \ref{fig:unique_usage_scores}, displays a box and whisker plot of the hashtag user scores for each of the cluster for each of the time periods.

\begin{figure}[ht]
    \centering
    \includegraphics[width=0.9\textwidth]{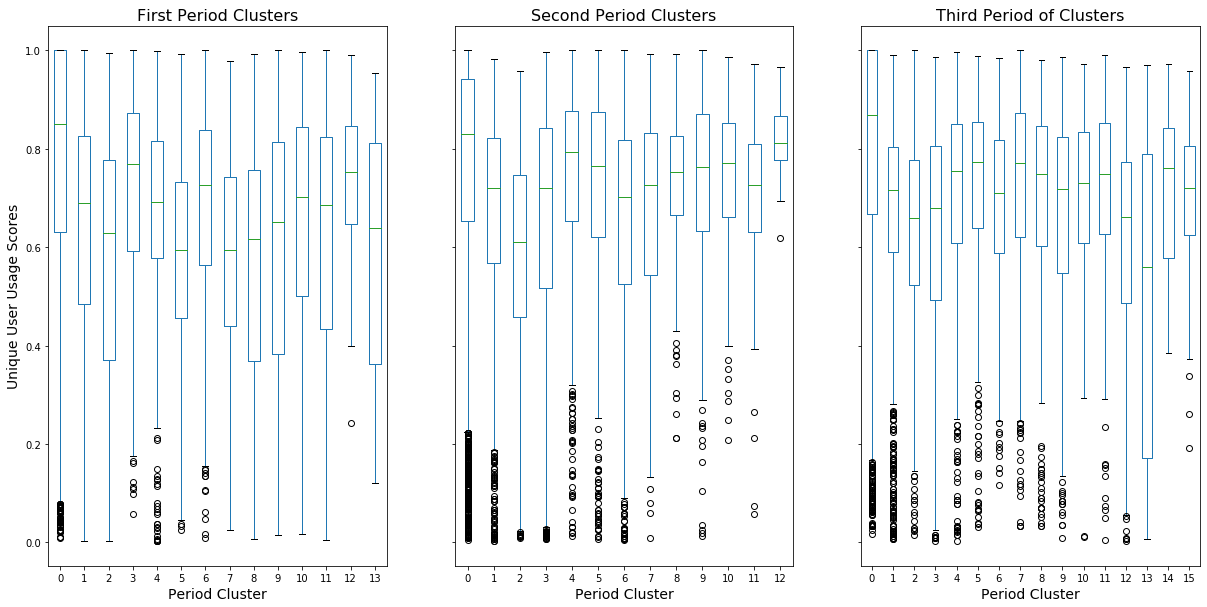}
    \caption[Top user scores for each hashtag in each cluster]{Box and whisker plot of the unique user scores for each of the hashtags within a cluster. Generally, each cluster has a relatively wide spread of scores which indicates hashtags that have many unique users and those with very few being present in each cluster. Many clusters also contain low-scoring outliers which are hashtags that are used by very few or even just one user.}
    \label{fig:unique_usage_scores}
\end{figure}

The unique user scores for each of the hashtags within each of the clusters show distinct differences between the clusters within each period. First, it is worth noting that the spread of scores for the different clusters tend to be consistently wide; most clusters have scores ranging from $1.0$ to near $0.2$. There are notable exceptions to this, like period 2 cluster 12 which focuses on online education, and period 1 cluster 12 which is an Italian-language cluster. Second, while there is a wide range of scores for the hashtags within any given cluster, most clusters often have outlier hashtags with very low user scores. That is to say, that most clusters have at least one hashtag that is often tweeted, but only by a few or one user. So, while the clusters often present distinct themes in their hashtags, the usage of the hashtags within the clusters can vary considerably. Some hashtags have a much more diverse base of usage while some are promulgated by only a few users. This result once again suggests that clusters of hashtags form around topical groups and that all hashtags are not equal, at least in terms of usage, within a topical group.

Overall, the user base of the different hashtag clusters demonstrate that not only are the clusters very different in their user make up, but so to are the user bases for individual hashtags within the clusters. Certain topical areas of hashtag usage, and by extension, discussion, tend to be driven by a small number of users, whereas others have a much more diverse user base. Furthermore, within clusters of hashtags, hashtags can vary quite a lot with their user bases, with some hashtags being promulgated by very few, or even one, users. So, even within a certain topical area of hashtags, a small group of users will control the usage of certain hashtags and possibly parts of the discussion as well. 

\subsection*{Detailed Analyses of Particular Clusters}

In this section, we analyze a few of the clusters more deeply. In particular, we analyze the usage statistics of different hashtags within the clusters as well as the verbiage associated with those clusters. 

\subsubsection*{First Period, Chinese-focused Cluster}

The first cluster of interest is a Chinese-focused cluster from the first period. This cluster was composed almost entirely of hashtags relating to either China or Wuhan. It should also be noted that this time period had a hashtag that featured the term `china' or `wuhan' in every single cluster except the cluster focused around business and commerce (period 1, cluster 11) and the cluster focused around U.S. Politics (period 1 cluster 6); around 86\% of the clusters in the first period featured a hashtag with one of these terms. In contrast, 68\% of the clusters in the second period and 50\% of the clusters in the third period had a hashtag with one of these terms. The following table, Table \ref{tab:period_1_cluster_3_hashtags}, displays a sample of some of the hashtags present within the cluster.

{\footnotesize
\begin{longtable}{|ll|ll|ll|}
\hline
\multicolumn{6}{|l|}{Period 1, cluster 3: Chinese-Focused}                                                                                                                                                                                                                                                                                                         \\ \hline
Most Used Hashtags       & \begin{tabular}[c]{@{}l@{}}Number of\\ Uses\end{tabular} & \begin{tabular}[c]{@{}l@{}}Hashtag with Highest\\ Original User Ratio\end{tabular} & \begin{tabular}[c]{@{}l@{}}User\\ Ratio\end{tabular} & \begin{tabular}[c]{@{}l@{}}Hashtags with\\ Lowest User Ratio\end{tabular} & \begin{tabular}[c]{@{}l@{}}User\\ Ratio\end{tabular} \\ \hline
WuhanVirus               & 145252                                                   & WeStandWithHongKong                                                                & 1.000                                                & myedgeprop                                                                & 0.058                                                \\
HongKong                 & 116401                                                   & HKpolicestate                                                                      & 1.000                                                & hongkonggenocide                                                          & 0.108                                                \\
CCP                      & 52252                                                    & Catastrophy                                                                        & 1.000                                                & usdjpy                                                                    & 0.112                                                \\
Hubei                    & 46059                                                    & Darkness                                                                           & 1.000                                                & china\_is\_territorist                                                    & 0.122                                                \\
Chinese                  & 44336                                                    & ProtestArt                                                                         & 1.000                                                & TechJunkieNews                                                            & 0.161                                                \\
WuhanPneumonia           & 31435                                                    & HKexit                                                                             & 0.998                                                & EiSamay                                                                   & 0.165                                                \\
Coronarivus              & 31152                                                    & Hubie                                                                              & 0.995                                                & hongkongprotest                                                           & 0.176                                                \\
WuhanCoronavirusOutbreak & 26612                                                    & timelapse                                                                          & 0.985                                                & CaptainTripps                                                             & 0.196                                                \\
LiWenliang               & 26352                                                    & stayclam                                                                           & 0.982                                                & Análisis                                                                  & 0.202                                                \\
HongKongProtests         & 26324                                                    & PLAAF                                                                              & 0.982                                                & EnvironmentHealth                                                         & 0.207                                                \\ \hline
\caption[Top raked hashtags from period 1, cluster 3 which is Chinese-focused in its hashtags]{Top raked hashtags from period 1, cluster 3 which is Chinese-focused in its hashtags. Many of the hashtags from this cluster are critical of the Chinese government for either its response to recent protests in Hong Kong or for its response to the COVID-19 outbreak from Wuhan.}
\label{tab:period_1_cluster_3_hashtags}
\end{longtable}
}

In general, the hashtags within this cluster express negative sentiment towards China and the Chinese government. Some of the most used hashtags within the cluster express support for Hong Kong protests and symbols of frustration with the Chinese government, like `$\#$LiWenliang' \cite{BBC:2020}. This trend is further emphasized with both those hashtags which are used by a diverse user base (highest original user ratio) and those used by very few users. Overall, there is a mix of anti-Chinese hashtags and pro-Hong Kong and pro-Tibet hashtags present within the cluster. So, in the first period, which are the early days of the COVID-19 pandemic there is a distinct vein of discussion within the COVID-19 discussions that is expressing negative sentiment toward the Chinese and Chinese government. It is interesting to observe that this negative sentiment is not limited to COVID-19 but also encompasses other issues with the Chinese government to include protests in Hong Kong. 

To get a better idea of nature of the discussion that uses these hashtags, we now turn to the text view and the words that co-occur with the hashtags in the cluster. The following figure, Figure \ref{fig:words_for_period_1_cluster_3} displays a word map of the common words that co-occur with hashtags from the cluster.

\begin{figure}[ht]
    \centering
    \includegraphics[width=0.9\textwidth]{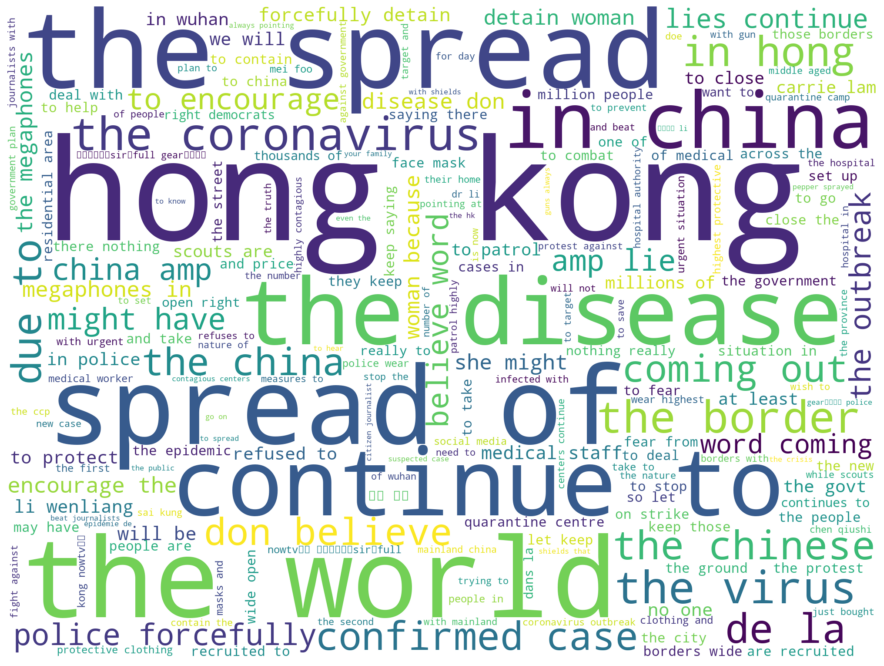}
    \caption[Word map of Chinese-focused cluster in the first period]{Word map of frequently occurring words and phrases that co-occur with hashtags from the Chinese-focused cluster in the first period. There is a significant amount of verbiage about the spread of the virus and about protests in Hong Kong.}
    \label{fig:words_for_period_1_cluster_3}
\end{figure}

As with the hashtags themselves, the words that often co-occur with the hashtags indicate a focus on the spread of the Coronavirus and Hong Kong related issues. Thus, it would seem that in the early period of the data there is a significant amount of negative discussion surrounding China and the Chinese government both for recent actions in Hong Kong and for being the source of the COVID-19 pandemic. When this result is combined with the user base analysis results of this cluster which indicate that this cluster has the most overlap between the users of its hashtags, it would seem that the discussion and condemnation of Chinese government actions during the early part of the pandemic is driven by relatively few users. This result also indicates that those users who were already critical of the either the Chinese response to Hong Kong protests or to the initial Chinese government reaction to the COVID-19 pandemic may be using the other calamity to draw attention to their calamity of focus. So, while Chinese related terms feature prominently in many of the clusters of hashtags in this period, there is also a distinct cluster of hashtag usage that is critical of the Chinese government.

\subsubsection*{First Period, Syrian Civil War Cluster}

The next cluster of interest is a cluster that features content centered on the Syrian Civil War. At first glance, this is already a strange cluster to have in a data set that was collected based on COVID-19 tweets; there is not an obvious connection between the two entities beside the fact that they are both significant, contemporary calamities. Additionally, the cluster has a high user ratio score, meaning different user accounts are using different hashtags. So, it would seem there is actually a diverse base of users supporting the different hashtags, but the content is almost solely focused on the Syrian Civil War. The following table, Table \ref{tab:period_1_cluster_13_hashtags}, displays some of the salient hashtags from the cluster.

{\footnotesize
\begin{longtable}{|ll|ll|ll|}
\hline
\multicolumn{6}{|l|}{Period 1, cluster 13: Syrian Civil War}                                                                                                                                                                                                                                                                                                 \\ \hline
Most Used Hashtags & \begin{tabular}[c]{@{}l@{}}Number of\\ Uses\end{tabular} & \begin{tabular}[c]{@{}l@{}}Hashtag with Highest\\ Original User Ratio\end{tabular} & \begin{tabular}[c]{@{}l@{}}User\\ Ratio\end{tabular} & \begin{tabular}[c]{@{}l@{}}Hashtags with\\ Lowest User Ratio\end{tabular} & \begin{tabular}[c]{@{}l@{}}User\\ Ratio\end{tabular} \\ \hline
CORONAVIRUS        & 38642                                                    & BILLGATES                                                                          & 0.955                                                & AssadGenocide                                                             & 0.122                                                \\
CHINA              & 10509                                                    & Nation                                                                             & 0.944                                                & Assad\_Torture                                                            & 0.123                                                \\
WUHAN              & 2397                                                     & ACTUALIZACIÓN                                                                      & 0.929                                                & Chemical\_Assad                                                           & 0.123                                                \\
NCOV19             & 1543                                                     & LAMORGESE                                                                          & 0.927                                                & TheResistance1776                                                         & 0.123                                                \\
IndianArmy         & 1316                                                     & FAKENEWS                                                                           & 0.908                                                & AssadCrimes                                                               & 0.123                                                \\
ALERT              & 1294                                                     & CINA                                                                               & 0.899                                                & PutinAtWar                                                                & 0.124                                                \\
BIOWEAPON          & 992                                                      & SINGAPORE                                                                          & 0.896                                                & WhiteHelmets                                                              & 0.125                                                \\
Syrie              & 981                                                      & BIOWEAPON                                                                          & 0.860                                                & InfoWars                                                                  & 0.215                                                \\
VIRUS              & 910                                                      & BEIJING                                                                            & 0.859                                                & TBT                                                                       & 0.233                                                \\
AssadGenocide      & 831                                                      & ALERT                                                                              & 0.850                                                & VIRUSCORONA                                                               & 0.241                                                \\ \hline
\caption[Hashtags from period 1 cluster 13, which has content and some hashtags devoted to the Syrian Civil War]{Hashtags from period 1 cluster 13, which has content and some hashtags devoted to the Syrian Civil War. Many of the hashtags within the cluster have non-overlapping users and are often all caps versions of more well known hashtags.}
\label{tab:period_1_cluster_13_hashtags}
\end{longtable}
}

\pagebreak
The hashtags present in this cluster differ from hashtags in other clusters. For one, many of the hashtags in use are all caps versions of other hashtags, such as `$\#$CORONAVIRUS' for `$\#$coronavirus'. Second, those hashtags which have very little overlap are almost all of the all caps variety while those with much less user overlap are more widely used and recognized hashtags, like `$\#$WhiteHelmets'. To get a better idea of the usage of the hashtags present in this cluster, the text co-occurring with the hashtags can be analyzed. The following figure, Figure \ref{fig:words_for_period_1_cluster_13}, displays a word map for the commonly used text that co-occurs with the hashtags in this cluster.

\begin{figure}[ht]
    \centering
    \includegraphics[width=0.9\textwidth]{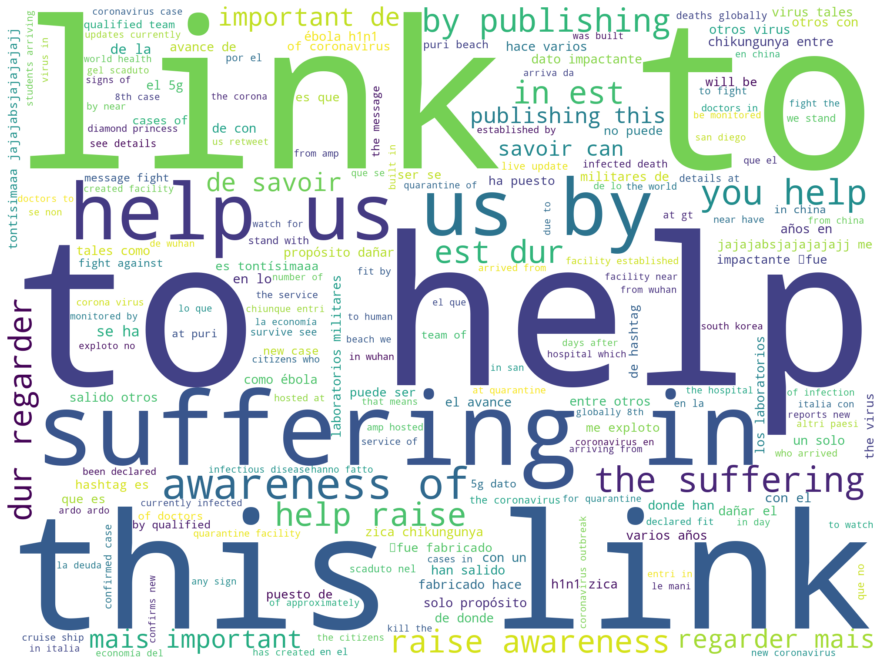}
    \caption[Word map of period 1, cluster 13, Syrian Civil War-focused cluster]{Word map of commonly occurring phrases and words from period 1, cluster 13. Much of the verbiage seems aimed at directing users to links to support various parties involved in the Syrian Civil War.}
    \label{fig:words_for_period_1_cluster_13}
\end{figure}

As with the hashtag themselves, the text co-occuring with the hashtags is different from the other clusters. Phrases like `link to' and `to help' feature prominently in the accompanying text and the accompanying text is much more focused in the primary topical area of the cluster than other clusters. Additionally, the text is multi-lingual in that there are mostly English words, but also words from Spanish, French, and others. So, from the text, these hashtags are used to promote awareness of the Syrian Civil War and to draw users to websites to provide support to various organizations involved in the Syrian Civil War. Thus, these hashtags seem to be aimed at using the COVID-19 pandemic to draw user interest to the Syrian Civil War. Additionally, it is also interesting to note that while the text is very uniform in its drawing attention to the Syrian Civil War, the user base is actually very diverse. This cluster has the most diverse user base of any cluster in the first time period. So, these results would suggest that while there is a diverse user base for this cluster, that this diversity is artificial; the users that use the hashtags within this cluster are probably related and possibly being centrally coordinated. At any rate, it is clear from the results that the hashtag usage in this cluster is meant to use the COVID-19 pandemic to draw attention to an entirely different calamity.

\subsubsection*{Second Period, Online Education Cluster}

Another cluster of interest is a cluster of hashtags that only exists in the second time period and focuses exclusively on online education-related hashtags. This cluster contains relatively few hashtags (37 in total) that all relate to online learning. It also exists only in the second period when most of the world entered some form of lock-down to slow the spread of the coronavirus. Despite the few number of hashtags in the cluster, this cluster has the most diverse user base in the second period and has all of its hashtags generally having a diverse usage (the lowest user score for a hashtag in the cluster is 0.619). The following table, Table \ref{tab:period_2_cluster_13_hashtags}, displays the some of the salient hashtags from the cluster.

{\footnotesize
\begin{longtable}{|ll|ll|ll|}
\hline
\multicolumn{6}{|l|}{Period 2, cluster 13: Online Education}                                                                                                                                                                                                                                                                                                 \\ \hline
Most Used Hashtags & \begin{tabular}[c]{@{}l@{}}Number of\\ Uses\end{tabular} & \begin{tabular}[c]{@{}l@{}}Hashtag with Highest\\ Original User Ratio\end{tabular} & \begin{tabular}[c]{@{}l@{}}User\\ Ratio\end{tabular} & \begin{tabular}[c]{@{}l@{}}Hashtags with\\ Lowest User Ratio\end{tabular} & \begin{tabular}[c]{@{}l@{}}User\\ Ratio\end{tabular} \\ \hline
education          & 9817                                                     & child                                                                              & 0.967                                                & Education                                                                 & 0.619                                                \\
onlinelearning     & 4386                                                     & teaching                                                                           & 0.907                                                & Learning                                                                  & 0.694                                                \\
edtech             & 4673                                                     & AcademicChatter                                                                    & 0.907                                                & STEMeducation                                                             & 0.712                                                \\
college            & 3530                                                     & college                                                                            & 0.901                                                & EdChat                                                                    & 0.714                                                \\
AcademicTwitter    & 3606                                                     & student                                                                            & 0.899                                                & intled                                                                    & 0.734                                                \\
distancelearning   & 3505                                                     & virtuallearning                                                                    & 0.886                                                & highered                                                                  & 0.739                                                \\
AcademicChatter    & 3204                                                     & universities                                                                       & 0.879                                                & edtech                                                                    & 0.751                                                \\
edchat             & 3594                                                     & AcademicTwitter                                                                    & 0.876                                                & HigherEd                                                                  & 0.761                                                \\
online             & 3235                                                     & students                                                                           & 0.874                                                & university                                                                & 0.774                                                \\
STEM               & 2936                                                     & distancelearning                                                                   & 0.867                                                & education                                                                 & 0.777                                                \\ \hline
\caption[Hashtags from period two cluster 12 which focus on online education]{Hashtags from period two cluster 12 which focus on online education. The hashtags used in this cluster have a diverse user base and are focused in that there are no hashtags that are not easily identifiable as being education-related in the cluster.}
\label{tab:period_2_cluster_13_hashtags}
\end{longtable}
}

As was mentioned previously, all of the hashtags in this cluster relate to education, especially online education, and have diverse user bases. In order to better understand the nature of the usage of the hashtags we turn to the words that co-occur with these hashtags. The following figure, Figure \ref{fig:words_for_period_2_cluster_12}, displays a word map for the frequently used words and phrases from the cluster.

\begin{figure}[ht]
    \centering
    \includegraphics[width=0.9\textwidth]{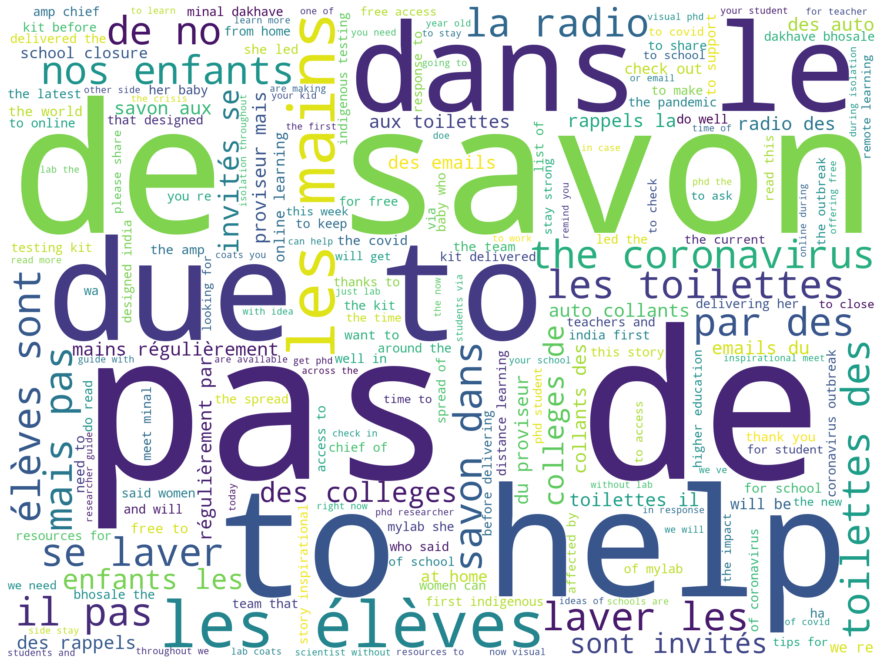}
    \caption[Word map for cluster 12 from period 2, which has education hashtags]{Word map for commonly used words and phrases present in cluster 12 from period 2, which has education hashtags. The verbiage is a mix of primarily English and French and has many phrases relating to school closures and new school hygiene policies.}
    \label{fig:words_for_period_2_cluster_12}
\end{figure}

Despite the hashtags all being in English, there is a surprising amount foreign language text, especially French, that co-occurs with these hashtags. Much of the text relates to the closing of schools, and new health-related procedures for schools. These hashtags were used not only to promote online learning solutions and products but also to advertise moving of courses to online and new school-related health procedures. So, this cluster highlights the use of hashtags in order to inform groups of users about both relevant happenings as well as education alternatives during a pandemic. In that sense, this cluster differs from many of the other clusters in that it focuses not only on different content but also seems more aimed at a sharing of non-partisan information to users.

\subsubsection*{Comparison of U.S. Politics Focused Clusters}

Finally, the U.S. politics focused clusters from periods one and three are analyzed to both understand their content and how the discussion topic of U.S. politics has changed over the course of the COVID-19 pandemic. In the first period, the U.S. politics focused cluster is slightly above average in terms of the user base diversity and has a relatively large number of hashtags at 343. The following table, Table \ref{tab:period_1_cluster_6_hashtags} displays the salient hashtags from the cluster.

{\footnotesize
\begin{longtable}{|ll|ll|ll|}
\hline
\multicolumn{6}{|l|}{Period 1, cluster 6: U.S. Politics}                                                                                                                                                                                                                                                                                                     \\ \hline
Most Used Hashtags & \begin{tabular}[c]{@{}l@{}}Number of\\ Uses\end{tabular} & \begin{tabular}[c]{@{}l@{}}Hashtag with Highest\\ Original User Ratio\end{tabular} & \begin{tabular}[c]{@{}l@{}}User\\ Ratio\end{tabular} & \begin{tabular}[c]{@{}l@{}}Hashtags with\\ Lowest User Ratio\end{tabular} & \begin{tabular}[c]{@{}l@{}}User\\ Ratio\end{tabular} \\ \hline
MAGA               & 13842                                                    & OpenBorders                                                                        & 1.000                                                & hillaryemails                                                             & 0.008                                                \\
Trump              & 11976                                                    & Newyork                                                                            & 0.999                                                & HAction                                                                   & 0.018                                                \\
FakeNews           & 9905                                                     & TheGreatAwakeing                                                                   & 0.994                                                & StopTheMadness                                                            & 0.048                                                \\
AmericaFirst       & 9150                                                     & DemCast                                                                            & 0.985                                                & bluelivesmatter                                                           & 0.062                                                \\
QAnon              & 8456                                                     & ThesePeopleAreSick                                                                 & 0.982                                                & ImpeachTrump                                                              & 0.105                                                \\
GatesFoundation    & 7768                                                     & GatesFoundation                                                                    & 0.981                                                & TRoom                                                                     & 0.106                                                \\
Dobbs              & 7752                                                     & IngrahamAngle                                                                      & 0.980                                                & ABQ                                                                       & 0.135                                                \\
Newyork            & 7131                                                     & TrustThePlan                                                                       & 0.960                                                & rockoftalk                                                                & 0.135                                                \\
FoxNews            & 6209                                                     & VoteBlueToEndThisNightmare                                                         & 0.959                                                & NM                                                                        & 0.140                                                \\
FreeZeroHedge      & 5736                                                     & JoeBiden                                                                           & 0.955                                                & Galaxy                                                                    & 0.148                                                \\ \hline
\caption[Hashtags from period one cluster 6 which focuses on U.S. Politics related clusters]{Hashtags from period one cluster 6 which focuses on U.S. Politics related clusters. There is a mix of hashtags associated with political news, political personalities and prominent politically-based conspiracy theories. The hashtags from this cluster also have a wide range of user bases in terms of the uniqueness of the users that use the hashtags}
\label{tab:period_1_cluster_6_hashtags}
\end{longtable}
}

\pagebreak
The hashtags are a mix of political commentary, hashtags related to prominent political figures, and hashtags typically related to conspiracy theories (i.e. `$\#$QAnon'). Some of the most used hashtags within the cluster relate directly to current U.S. President Donald Trump (i.e. `$\#$MAGA' and `$\#$Trump') while some of the most diverse hashtags in the cluster are anti-President Trump (i.e. `$\#$VoteBlueToEndThisNightmare' and `$\#$DemCast'). It is also interesting to note that the two hashtags with the least diverse user base --- which have scores well outside the inter-quartile range of user scores for the cluster --- have only one user who posts both hashtags, 119 and 56 times respectively. So, the cluster contains various hashtags related to various elements of U.S. Politics, including those which are generally associated with partisan content. Thus, in some respects, as with the Syrian War Cluster this cluster contains hashtags which are attempting to use the COVID-19 pandemic in order to draw attention to certain political views or ideas. 

To get a better idea of the hashtag usage in the first period's U.S. Politics cluster, I analyzed the text which co-occurs with the hashtags. As with the previous analyses of other clusters, the following figure, Figure \ref{fig:words_for_period_1_cluster_6}, displays the word map for the co-occurring text.

\begin{figure}[ht]
    \centering
    \includegraphics[width=0.9\textwidth]{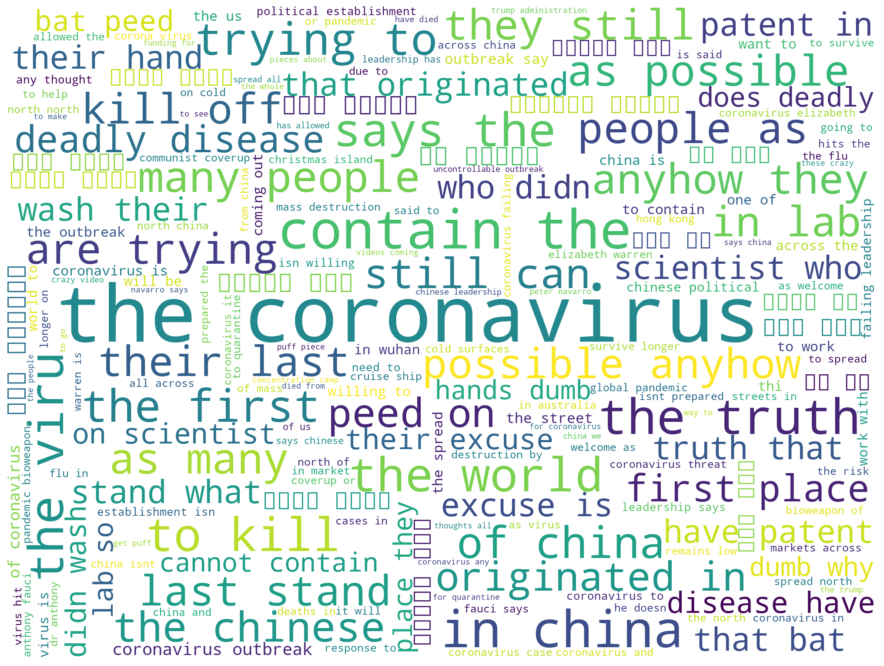}
    \caption[Word map of period 1, cluster 6, which is composed of U.S. politics related hashtags]{Word map of commonly occurring phrases and text that co-occur with hashtags from period 1, cluster 6, which is composed of U.S. politics related hashtags. There is a wide range of verbiage employed in the cluster, with much of it focusing on the COVID-19 pandemics origins in China and its subsequent spread.}
    \label{fig:words_for_period_1_cluster_6}
\end{figure}

\pagebreak
Generally, no phrase or word is particularly dominant in this cluster except for `the coronavirus' itself. Much of the text mentions the origin of the pandemic in China as well as the global spread of the virus. In this first period U.S. politics cluster, much of the verbiage centers around the origin of the virus and its possible effects. So, much of the discussion from politically-connected hashtags during the first period focuses on China's role in the origin and spread of the COVID-19 pandemic.  

Turning to the third period U.S. politics focused cluster, there are similar usage patterns among the hashtags. The following table, Table \ref{tab:period_3_cluster_5_hashtags}, displays the salient hashtags from the third period U.S. politics cluster.

{\footnotesize
\begin{longtable}{|ll|ll|ll|}
\hline
\multicolumn{6}{|l|}{Period 3, cluster 5: U.S. Politics}                                                                                                                                                                                                                                                                                                     \\ \hline
Most Used Hashtags & \begin{tabular}[c]{@{}l@{}}Number of\\ Uses\end{tabular} & \begin{tabular}[c]{@{}l@{}}Hashtag with Highest\\ Original User Ratio\end{tabular} & \begin{tabular}[c]{@{}l@{}}User\\ Ratio\end{tabular} & \begin{tabular}[c]{@{}l@{}}Hashtags with\\ Lowest User Ratio\end{tabular} & \begin{tabular}[c]{@{}l@{}}User\\ Ratio\end{tabular} \\ \hline
Trump              & 189109                                                   & Satanism                                                                           & 0.988                                                & hillaryemails                                                             & 0.030                                                \\
FakeNews           & 74076                                                    & DeutscheBank                                                                       & 0.987                                                & PoliticalViews                                                            & 0.037                                                \\
MAGA               & 70999                                                    & Morons                                                                             & 0.986                                                & HAction                                                                   & 0.040                                                \\
FoxNews            & 67531                                                    & Socialists                                                                         & 0.981                                                & drudge                                                                    & 0.051                                                \\
WWG1WGA            & 49348                                                    & ShutItDown                                                                         & 0.978                                                & slate                                                                     & 0.064                                                \\
KAG                & 48339                                                    & AlexJones                                                                          & 0.977                                                & newsabq                                                                   & 0.065                                                \\
Trump2020          & 46790                                                    & 2ndAmendment                                                                       & 0.971                                                & abqfm                                                                     & 0.067                                                \\
OneVoice1          & 43457                                                    & hypocrisy                                                                          & 0.969                                                & rockoftalk                                                                & 0.076                                                \\
QAnon              & 41446                                                    & DrainingTheSwamp                                                                   & 0.960                                                & NewsVideo                                                                 & 0.079                                                \\
CoronavirusUSA     & 34819                                                    & Bullshit                                                                           & 0.960                                                & bluelivesmatter                                                           & 0.083                                                \\ \hline
\caption[Important hashtags from period 3 cluster 5 which has U.S. Politics related hashtags]{Important hashtags from period 3 cluster 5 which has U.S. Politics related hashtags. As with other U.S. politics clusters, this cluster is a mix of hashtags from political news sources, political personalities, and politically-motivated conspiracy theories. Relative to the first period's U.S. Politics cluster, there is an increase in the use of more politically inflammatory hashtags.}
\label{tab:period_3_cluster_5_hashtags}
\end{longtable}
}

As with the first period's U.S. politics cluster, many of the hashtags surround political commentary, politically-motivated conspiracy theories, and high profile politicians. Unlike the first period, however, the hashtags with the lowest user scores have larger users bases (i.e. 1 unique user versus 7 for the lowest scoring hashtag). There is also an increase in the number of hashtags being used, 343 in period one versus 610 in period three. To get a better sense of the differences between the two periods, the verbiage of the text co-occurring with the hashtags was then analyzed. The following figure, Figure \ref{fig:words_for_period_3_cluster_5}, displays a word map of the commonly used phrases and words from the cluster

\pagebreak
\begin{figure}[ht]
    \centering
    \includegraphics[width=0.9\textwidth]{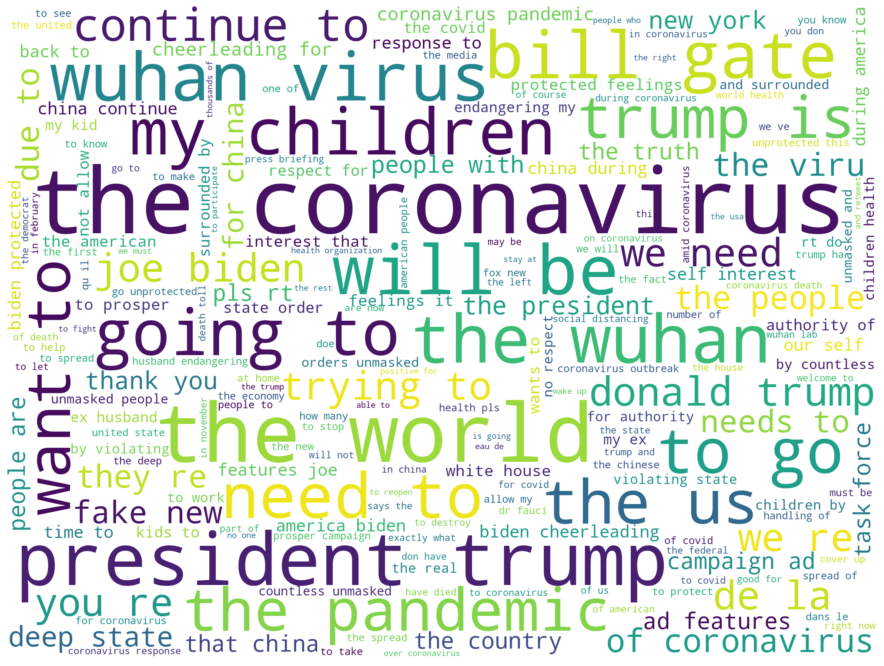}
    \caption[Word map of period 3 cluster 5, which consists of U.S. politics related hashtags]{Word map of the commonly used phrases and words from period 3 cluster 5, which consists of U.S. politics related hashtags. Much of the verbiage focuses on personalities, like President Donald Trump or Bill Gates who has been erroneously linked to the COVID-19 outbreak by conspiracy theories.}
    \label{fig:words_for_period_3_cluster_5}
\end{figure}

In this time period phrases surrounding personalities become much more used, most especially President Donald Trump. Other personalities like Bill Gates and Joe Biden are also frequently mentioned in the political hashtag tweets. So, from the first period to the third period, much of the verbiage shifts from a focus on the virus' origins in China and its spread, to personalities who are often political in nature or who have been linked to the coronavirus, even if erroneously, such as Bill Gates. So, overall there is a shift in the focus of the verbiage paired with the U.S. politics hashtags over the course of the pandemic. This shift generally goes from an external focus to an internal focus relative to the United States.

Finally, to get a more complete sense of the differences between the period 1 and period 3 clusters, we directly compare the membership of these two clusters. The user base overlap between the two time periods is small; only 8\% of the union of the two time periods' users are present in both time periods. These users are generally high profile users like politicians or news sources. There is also a 31.8\% overlap in the hashtags used between the two time periods. To get a sense of the differences in membership between these clusters, we analyzed the most important hashtags that exist in only one of the two clusters. It should be noted that many of the hashtags that exist in only one cluster do exist in the other time period that their cluster does not belong to, but not within the U.S. politics cluster of that time period. The following table, Table \ref{tab:comparison_between_top_used_politics_hashtags}, displays a side-by-side comparison of the hashtags between the two clusters.

\pagebreak
{\footnotesize
\begin{longtable}{|llll|llll|}
\hline
\begin{tabular}[c]{@{}l@{}}Period 1 Only\\ Hashtags\end{tabular} & \begin{tabular}[c]{@{}l@{}}Number of\\ Uses\end{tabular} & \begin{tabular}[c]{@{}l@{}}Period 3 Only\\ Hashtags\end{tabular} & \begin{tabular}[c]{@{}l@{}}Number of\\ Uses\end{tabular} & \begin{tabular}[c]{@{}l@{}}Period 1 Only\\ Hashtags\end{tabular} & \begin{tabular}[c]{@{}l@{}}User\\ Ratio\end{tabular} & \begin{tabular}[c]{@{}l@{}}Period 3 Only\\ Hashtags\end{tabular} & \begin{tabular}[c]{@{}l@{}}User\\ Ratio\end{tabular} \\ \hline
coronavirusaustralia                                             & 35561                                                    & smartnews                                                        & 28748                                                    & openborders                                                      & 1.000                                                & morons                                                           & 0.986                                                \\
newyork                                                          & 7131                                                     & 5g                                                               & 27165                                                    & newyork                                                          & 0.999                                                & socialists                                                       & 0.981                                                \\
racism                                                           & 5016                                                     & america                                                          & 18568                                                    & thegreatawakeing                                                 & 0.994                                                & shutitdown                                                       & 0.978                                                \\
censorship                                                       & 2426                                                     & oann                                                             & 16456                                                    & bias                                                             & 0.946                                                & 2ndamendment                                                     & 0.971                                                \\
thegreatawakeing                                                 & 2017                                                     & pressbriefing                                                    & 11899                                                    & virginia                                                         & 0.943                                                & hypocrisy                                                        & 0.969                                                \\
coronovirus                                                      & 1829                                                     & new                                                              & 8650                                                     & trumpbudget                                                      & 0.939                                                & justasking                                                       & 0.959                                                \\
trumpbudget                                                      & 1689                                                     & nyt                                                              & 8276                                                     & chaos                                                            & 0.939                                                & nyt                                                              & 0.959                                                \\
democracy                                                        & 1613                                                     & deepstate                                                        & 7982                                                     & confirms                                                         & 0.926                                                & senatorforsale                                                   & 0.958                                                \\
zerohedge                                                        & 1574                                                     & senatorforsale                                                   & 7389                                                     & lnpfail                                                          & 0.923                                                & justsaying                                                       & 0.956                                                \\
iran                                                             & 1546                                                     & americans                                                        & 7113                                                     & earthquakes                                                      & 0.922                                                & antivaxx                                                         & 0.955                                                \\ \hline
\caption[Comparison of the important hashtags that either in the U.S. politics cluster in period one or period three, but not both]{Comparison of the important hashtags that either in the U.S. politics cluster in period one or period three, but not both. Generally, there is an increase in conspiracy-related hashtag and inflammatory hashtag usage from period 1 to period 3.}
\label{tab:comparison_between_top_used_politics_hashtags}
\end{longtable}
}

There are some distinct differences in those hashtags only used in one of the clusters and not in the other. First, the salient period one only hashtags feature Australian-related hashtags like `$\#$coronavirusaustralia' and `$\#$lnpfail' which are not in period three. Also, there is a rise in the use of conspiracy-related hashtags in the third period only hashtags, such as `$\#$5g', `$\#$deepstate', `$\#$antivaxx' and more inflammatory hashtags like `$\#$morons' or `$\#$senatorforsale'. So, there is not only a regional shift in terms of the difference in the U.S. politics hashtags between periods one and three, but also one toward more polarizing and contentious hashtags over time as well.

Overall, the cluster of U.S. politics-related hashtags differs over the course of the pandemic. The use of some hashtags (approximately a third) remains the same, but the verbiage associated with those hashtags changes from a spread of the disease and Chinese focus to a personality and conspiracy-theory focus. Additionally, the hashtags used in only one time period also show some distinct differences between the time periods. So, while an easily defined topical discussion characterized by the hashtags being used can be persistent over the course of the pandemic, the nature of that topical discussion cluster changes. It is also worth noting that known conspiracy theory related hashtags are always present in the U.S. politics cluster, which demonstrates an strong connection between the two over the course of the COVID-19 pandemic.

\section*{Discussion}

There are several findings from this study and results to inspire future research. First, through a scalable technique, like MVMC, it is possible to extend multi-view clustering to a task like clustering hashtags in large-scale social media data. Large scale social media data often requires clustering of tens or even hundreds of thousands of objects and the ability to handle partially incomplete data. The MVMC procedure can successfully deal with both of these conditions in the data and produce meaningful clusters. Use of the MVMC technique also found that certain views, in their current form of feature representation, like URLs which co-occur with hashtags in tweets, were not useful in finding a cluster structure in the hashtags. Also the use of multi-view clustering on hashtag can ameliorate problems with previous attempts to cluster hashtags based on just one view. For example, incorporating text and shared users can overcome the observed phenomenon where two hashtags are very related in usage, and should be clustered, but are never used by the same users. Thus, through a technique like MVMC it is possible to incorporate all of the previous research on clustering hashtags, that have used co-occurring text or users, into one cohesive model and clustering. 

Second, hashtag usage patterns during the COVID-19 pandemic displayed dynamic behavior at both the individual and cluster levels. While the data collected is certainly an incomplete picture of the discussions happening on twitter due to API restrictions and the terms used to create the data, it can still offer some insights. From the early days of public awareness about the the pandemic in February of 2020, there was an increase in the number of unique hashtags being used and the the number of unique users participating in the COVID-19 discussion on twitter. The usage patterns of hashtags, however, varied over the course of the pandemic with there being an initially high rate of hashtag usage that drops when the number of unique users increase, and then increases again as the number of unique users levels of in late March/ early April. This suggests that when a major exogenous shock happens to social media users, like a pandemic, there will be an initial phase of interaction without hashtags, and then a move to start re-using hashtags, likely as a tool to aid in finding and participating in discussions.

At the cluster level, the data showed there were three main periods of clusters of hashtags present in the data. The daily hashtag clusterings could themselves be clustered into three distinct time periods of clusterings based solely on the pairwise similarity between the daily clusterings. In general, the cluster structure of hashtags went from a large number of small clusters to fewer, larger clusters and then back to smaller more numerous clusters. This macro temporal pattern in the cluster structure mirrors those findings from the use of individual hashtags and supports the conclusions that there was a surge in COVID-19 twitter discussion which produced an intermediary period of hashtag usage which then settled into a new pattern of hashtag usage different from what was observed prior to the user surge. 

Using the knowledge that the daily hashtag clustering breaks into three periods, we then created an ensemble clustering for each of these periods. This ensemble clustering allows for a clustering analysis of a prototypical clustering for the entire time period. The results of the analyses of these ensembled clusterings produced some interesting insights into the nature of some of the topical discussions happening during the pandemic. Firstly, there are some topics which have been persistent over the course of the pandemic, like commerce, the economy, U.S. politics, and news. Other topical groups like online education or negative-sentiment discussion about the Chinese government are more transitory over the course of the epidemic. From these clusters it was also observed that some topical groups are intending to direct COVID-19 discussion to other topics like the Syrian Civil War or protests in Hong Kong. So, it would seem from the nature of the clusters present that there is the presence and use of hashtags that are meant to use the COVID-19 pandemic to draw attention to other causes or ideas. So, hashtags can not only be a means helping users to find and participate in discussions but also as means of shaping user engagement and the discussions themselves.

From the results presented in this work there are several avenues for future research. First, this study focused on the use of hashtags in order to understand topical discussions taking place, which naturally discounts users who do not use hashtags. As was seen in the data section, there is a sizeable amount of the population of users that are posting content related to the COVID-19 discussions that do not use hashtags. So, a future area of research would be to look more broadly at the concepts that users are employing in their tweets. So, instead of just clustering on hashtags, one could look at clustering on hashtags and topical labels from something like the tweet text. Second, for the multi-view clustering of large scale clustering of hashtags in social media data there is a need for future research on the appropriate views and how to feature engineer those views to be useful. The URLs view of the data ended up being unhelpful for the found clusters, and this seems to be due in large part to the fact that there was very little overlap on exact URLs. So, there were situations in which essentially the same story or piece of news was used with two different hashtags in two different tweets, but because the URLs were not exactly the same, those hashtags were not recognized as being similar by the method. So, a means of processing the URLs to do something like just using the top level domains should be tried in future work. Additionally, previous research on misinformation during the COVID-19 pandemic has demonstrated that it can spread quickly by mechanisms like retweeting. It would be of value to create a tweet type view to characterize what type of tweets are being used with certain hashtags. Such a view may help with distinguishing between clusters of hashtags used for misinformation versus those used for more legitimate information. Second, there is also potential for future research in better cleaning and representing real-world data for multi-view clustering. For example, the tweets used in this study were not filtered by language. Performing a filtering step like only using English-language tweets could lead to more nuanced and meaningful clusters. Also, we adopted a heuristic graph learning procedure to form the view graphs for multi-view clustering of this data due to scalability issues with many of the more sophisticated graph learning procedures. Thus, an important avenue for future research is to find a graph learning procedure for MVMC that can better fit the intrinsic structure of the data, but that is also scalable to hundreds of thousands of entities. Finally, the data used in this study was only a sample of the twitter data pertaining to COVID-19. It would be interesting to see if different COVID-19 twitter data yield the same results and if there are differences between different social media platforms that also employ hashtags. 

%%%%%%%%%%%%%%%%%%%%%%%%%%%%%%%%%%%%%%%%%%%%%%
%%                                          %%
%% Backmatter begins here                   %%
%%                                          %%
%%%%%%%%%%%%%%%%%%%%%%%%%%%%%%%%%%%%%%%%%%%%%%

\begin{backmatter}

\section*{List of Abbreviations}
MVMC: Multi-view Modualrity Clustering, kNN: k-Nearest Neighbor, URL: Uniform Resource Locator

\section*{Declarations}
\section*{Availability of Data and Material}
The datasets used and/or analysed during the current study are available from the corresponding author, or CASOS institute at casos@cmu.edu on reasonable request. Python implementations of the methods presented in this study are available from the corresponding author on reasonable request.

\section*{Competing Interests}
The authors declare that they have no competing interests.

\section*{Ethics Approval and Consent to Participate}
Not applicable.

\section*{Funding}
This material is based upon work supported by the National Science Foundation Graduate Research Fellowship (DGE 1745016), Department of Defense Minerva Initiative (N00014-15-1-2797), and Office of Naval Research Multidisciplinary University Research Initiative (N00014-17-1-2675). Any opinion, findings, and conclusions or recommendations expressed in this material are those of the authors and do not necessarily reflect the views of the National Science Foundation, Department of Defense, or the Office of Naval Research.

\section*{Author's Contributions}
Iain Cruickshank and Kathleen M. Carley have developed together the theoretical setup of the study. Iain Cruickshank created the algorithmic framework, conducted the quantitative analyses and wrote the paper. Kathleen M. Carley has supervised the entire project. 

%%  After submission of the .TEX file,                     %%
%%  you will be prompted to submit your .BBL file.         %%

% if your bibliography is in bibtex format, use those commands:
\bibliographystyle{bmc-mathphys} % Style BST file (bmc-mathphys, vancouver, spbasic).
\bibliography{main}      % Bibliography file (usually '*.bib' )

\end{backmatter}
\end{document}